\documentclass[preprint,12pt]{elsarticle}

\usepackage{amssymb}
\usepackage{amsthm}

\usepackage{amsmath} 

\usepackage{tikz}
\usetikzlibrary{arrows, arrows.meta} 
\usetikzlibrary{decorations.pathreplacing}
\usetikzlibrary{decorations.pathmorphing,patterns}
\usetikzlibrary{calc,patterns,decorations.markings}
\usetikzlibrary{positioning}
\usepackage{graphicx}
\usepackage{caption}
\usepackage{changepage}
\usepackage{adjustbox}
\usepackage{lipsum}
\usepackage{subcaption}

\usepackage{calrsfs}
\usepackage[showframe=false]{geometry}
\DeclareMathAlphabet{\pazocal}{OMS}{zplm}{m}{n}

\usepackage[section]{placeins}

\theoremstyle{definition}

\theoremstyle{remark}

\graphicspath{{Figs/}}

\usepackage{float}
\newcommand{\etal}{\textit{et al}. }

\def\etc.{etc,.\spacefactor=\the\sfcode`\c}

\begin{document}
\begin{frontmatter}

\title{Central algorithms for accurately predicting non classical non-linear waves in Dense Gases over simple geometries
}
\author[1,2]{Ramesh Kolluru\corref{cor1}}
\ead{rameshkolluru43@gmail.com,kollurur@iisc.ac.in}
\cortext[cor1]{Corresponding author} 
\author[1]{S. V. Raghurama Rao}
\ead{raghu@iisc.ac.in}
\ead{svraghuramarao@gmail.com}
\author[3]{G.N.Sekhar}
\ead{gnsbms@gmail.com}

\address[1]{Department of Aerospace Engineering, Indian Institute of Science, Bangalore\vspace{0.1cm}}
\address[2]{Department of Mechanical Engineering, B.M.S.College of Engineering, Bangalore\vspace{0.1cm}}
\address[3]{Department of Mathematics, B.M.S.College of Engineering, Bangalore\vspace{0.1cm}}

\begin{abstract}
Non-classical non-linear waves exist in dense gases for large specific heats at pressures and temperatures of the order of critical point values. These waves behave precisely opposite to the classical non-linear waves, with inverted classical waves like the expansion shocks which do not violate entropy conditions.  More complex equation of state (EOS) other than the ideal or perfect EOS is typically used in describing dense gases. Algorithm development with non-ideal/real gas EOS and application to dense gasses is gaining importance from a numerical perspective. Extending the algorithms designed for perfect gas EOS to dense gas flows with arbitrary real gas EOS is non-trivial. Most of the algorithms designed for prefect gas EOS are modified significantly when applied to real gas EOS. These algorithms can become complicated and some times impossible based on the EOS under consideration. The objective of the present work is to develop central solvers with smart diffusion capabilities independent of the eigenstructure and extendable to any arbitrary EOS. Euler equations with van der Waals EOS along with two newly developed algorithms, Method of Optimal Viscosity for Enhanced resolution of shocks (MOVERS+) and Riemann Invariants based Contact capturing Algorithm (RICCA), are used to simulate dense gasses over simple geometries.  Various  One Dimensional (1D) and Two Dimensional (2D) benchmark test cases are validated using these algorithms, and the results are compared with the those obtained from the literature.
\end{abstract} 
\end{frontmatter}


\section{Introduction}
\label{intro}
Occurrence of classical nonlinear waves like shocks and expansion fans are common phenomena that are observed in high speed flow of a compressible gas. Most of the analysis is carried out assuming the gas to follow perfect/ideal gas equation of state. In this context, expansion shocks and compression fans are not valid solutions as they violate the entropy condition. 

The behaviour of high-speed flow in the dense regime ({\em i.e.}, at conditions close to thermodynamic critical point) is gaining attention from both experimental as well as numerical perspective. In the dense gas region, perfect gas EOS is not valid and real gas effects play a crucial role in predicting the dynamic behaviour of the gases.

\begin{figure}[htb!] 
\begin{center}
 \resizebox{80mm}{!}{\includegraphics{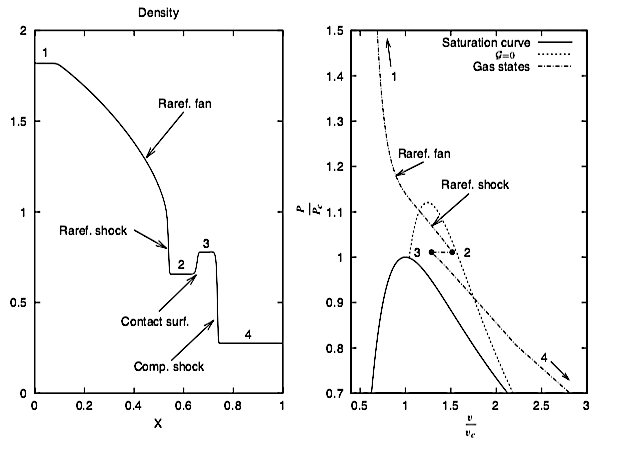}}
\end{center}
\caption{ Non-Classical Waves (Shock formation in the expansion region or during expansion process): Picture Courtesy \cite{alberto}. In this figure, $\mathcal{G}$ represents the fundamental derivative.}   
\label{ncw}
\end{figure}

Dense gases are usually \lq\textit{single-phase vapours whose thermodynamic state is close to saturation conditions or thermodynamical critical point}\rq.  These gases exhibit non-classical behaviour ({\em i.e.}, occurrences of non-classical waves).  \emph{Non-classical waves are the waves where mixed compression-expansion fans and expansion shocks occur without violating the entropy conditions (or the second law of thermodynamics)}, as shown in figure (\ref{ncw}). 

Examples of dense gases are BZT (Bethe-Zel'dovich-Thompson) fluids, refrigerants, hydrocarbons, perfluorocarbons or siloxanes and heavy polyatomic fluids which are commonly used in engineering applications as heat transfer fluids. These fluids have many practical applications, for example in energy-conversion cycles operating on low-temperature heat sources, such as Organic Rankine Cycle (ORC) \cite{Nalferez} and heavy gas wind tunnels \cite{Brown}. 
%

The non classical wave behaviour in dense gases can be justified through the second law of thermodynamics and shock theory. The relationship between the entropy change $\Delta s$ and the specific volume $v$ across a weak shock is given by (\ref{entropy}) as described in references (\cite{borisov,cramer89,cramer91_1,cramer91_2,cramer84,cramer86_2,cramer86_1,cramer87,menikoff})
\begin{equation}
\label{entropy}
\Delta s = -\left( \frac{\partial^2 p}{\partial v^2}\right) \frac{\left(\Delta v\right)^3}{12 T}.
\end{equation}
The nonlinear dynamics of dense gases are governed by an important property called the \emph{fundamental derivative}, $\Gamma$, of gas dynamics \cite{thompson} as given by
 \begin{equation}
 \label{fd}
 \Gamma = \frac{v^3}{2a^2}\left(\frac{\partial^2 p}{\partial v^2}\right)_s.
\end{equation}
From \eqref{fd} it can be observed that the curvature of an isentrope is given by $\left( \frac{\partial^2 p}{\partial v^2}\right)$ which becomes zero at critical points. In the case of dilute or perfect gases away from the critical point as shown in figure (\ref{ncw}), the curvature of the isentrope is always positive $\left( \frac{\partial^2 p}{\partial v^2}\right) > 0$ which enforces $\Delta v < 0$ satisfying $\Delta s >0$ . For dense gases it is possible that the curvature of isentrope can be negative given by $\left( \frac{\partial^2 p}{\partial v^2}\right) <0$ and in order to have $\Delta s > 0$,  $\Delta v > 0 $ must be satisfied. This is the reason why the expansion wave steepens and compression wave spreads without violating the entropy \cite{Brown}. 

 The Fundamental derivative \eqref{fd} can also be interpreted as the rate of change of speed of sound with reference to density as given in (\ref{afd}) for an isentropic process \cite{Brown}.  
\begin{equation}
\Gamma = 1 + \frac{\rho}{a}\frac{\partial a}{\partial \rho}
\label{afd}
\end{equation}
 It can be observed from (\ref{afd}) that for a perfect gas EOS in an isentropic process ($p\propto \rho^{\gamma}$) and with $a^{2} = \frac{\gamma p}{\rho}$, the fundamental derivative becomes $\Gamma = \frac{1 + \gamma}{2}$. For perfect gas EOS, $\gamma >1$,  and therefore $\Gamma >1$ is always true. For Dense gases the value of fundamental derivative can be $\Gamma > 0$ or  $\Gamma < 0$, depending upon the EOS being considered. The existence of non-classical non-linear waves basically depends on the sign of this fundamental derivative.

Initially, shock tube studies were confined to gases that produce classical wave fields (regular shocks and expansion waves) where entropy conditions are satisfied by both shock and expansion fans. Any violation of entropy conditions were not considered as physical.  Bethe \cite{Bethe} was the first one to analyse the possibility of negative values of fundamental derivative, followed by Zel'dovich \cite{Zeldovich} and Thompson \cite{thompson}.  Borisov \etal \cite{borisov} were the first to demonstrate the non-classical behaviour of gas in a shock tube.  As per the literature, this was the first instance where a shock tube experiment was used to investigate the nonclassical behaviour. Since then several others authored and explored nonclassical dense gas dynamics, with particular attention to the creation and evolution of expansion shocks in the region of negative nonlinearity (\cite{borisov,cramer89,cramer91_1,cramer91_2,cramer84,cramer86_2,cramer86_1,cramer87,menikoff}). They have demonstrated that in dense gases expansion shocks and smooth variations in the regions of shocks are physically possible without violating the entropy. 
  
Cinnella \cite{cinnella} proposed a simplification of Roe's scheme to solve for dense gases. Though this simplification does not satisfy the usual Roe's conditions for the approximate Riemann solver exactly, it reduces the complexity and computational cost. Higher order extension of the schemes are carried out using MUSCL method. 
Congedo \etal \cite{Congedo} studied the dense gas behaviour in turbo-machinery. In this work HLL scheme was been used in evaluating fluxes at the interfaces and MUSCL type reconstruction was used for higher order accuracy. The gradients at cell centers were evaluated using least squares formula. 
Argrow  \cite{argrow} conducted a numerical simulation of dense gas flows using a TVD Predictor-Corrector scheme for 1D Euler equations with van der Waals EOS. 
Brown and Argrow \cite{Brown} used a predictor-corrector TVD scheme based on the Davis-Roe flux limited method to simulate the flow of non-classical dense gas over simple geometries. They used van der Waals EOS which is a representative equation for heavy fluorocarbons with high specific heat and conditions near thermodynamic critical point. Comparisons have been made with equivalent perfect gas EOS. 
  
  Most popular numerical methods developed for solving the hyperbolic systems of conservation laws, the Riemann solvers, are heavily tied to the eigen-structure of the underlying systems.  They need significant modifications for equations of state other than the ideal one. The motivation for this work is to present algorithms for solving dense gas flows which do not require any modifications for any complicated equations of state.  In the present work, algorithms having controllable numerical diffusion and independent of the eigenstructure are utilised in resolving the non-classical waves on simple geometries. A recently developed central solvers MOVERS+ and RICCA  \cite{Ramesh_FP}\cite{rkolluru} are used to solve Euler equations with van der Waals EOS. These two algorithms are designed to capture steady contact discontinuities exactly and are low in numerical diffusion otherwise. This paper is arranged as follows.  Section \ref{sec:GE} reviews the governing equations and the numerical method used in solving them. In sections(\ref{sec:RICCA} $ \& $ \ref{sec:movers+}) the basic idea of the algorithms are explained briefly. The results obtained for benchmark test cases in 1D and 2D are presented in section(\ref{sec:results}).
   
\section{Governing equations and Numerical Models} \label{sec:GE}

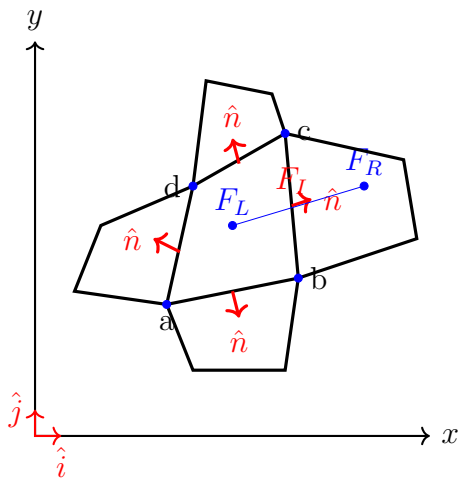
\begin{figure}[htb!]
\begin{center}
\begin{tikzpicture}[scale=1.75]
\draw [<->,thick] (0,3) node (yaxis) [above] {$y$} |- (3,0) node (xaxis) [right] {$x$};
\draw [<->,red,thick] (0,0.2) node (yaxis) [left] {$\hat{j}$} |- (0.2,0) node (xaxis) [below] {$\hat{i}$};
\draw [very thick] (1,1)node (xaxis) [below] {a} -- (2,1.2) --  (1.9,2.3) node (xaxis) [right] {c}-- (1.2,1.9)node (xaxis) [left] {d} -- (1,1) ;
\draw [->,red,very thick](1.5,1.1) -- (1.55,0.9) node (yaxis) [below] {$\hat{n}$ };
\draw [->,red,very thick](1.95,1.75) node (xaxis) [above] {$F_I $}-- (2.1,1.8) node (xaxis) [right] {$\hat{n} $};
\draw [->,red,very thick](1.55,2.07) -- (1.5,2.26) node (yaxis) [above] {$\hat{n}$};
\draw [->,red,very thick](1.1,1.4) -- (0.9,1.5) node (xaxis) [left] {$\hat{n} $};
\draw [blue,fill] (1.5,1.6) circle [radius=0.03] node (taxis) [above]{$F_L$}  ;

\draw [blue,very thin] (1.5,1.6) --  (2.5,1.9);
\draw [blue,fill] (2.5,1.9) circle [radius=0.03] node (taxis) [above]{$F_R$};

\draw [very thick] (2,1.2)  --  (2.9,1.5) -- (2.8,2.1) -- (1.9,2.3) ;
\draw [very thick](2,1.2)node (xaxis) [right] {b} --  (1.9,0.5) -- (1.2,0.5) -- (1,1) ;
\draw[very thick] (1,1) --  (0.3,1.1) -- (0.5,1.6) -- (1.2,1.9) ;
\draw [very thick](1.2,1.9) --  (1.3,2.7) -- (1.8,2.6) -- (1.9,2.3) ;
\draw [blue,fill] ( 1,1) circle [radius=0.03] ;
\draw [blue,fill] (2,1.2) circle [radius=0.03] ;
\draw [blue,fill] (1.9,2.3) circle [radius=0.03] ;
\draw [blue,fill] (1.2,1.9) circle [radius=0.03] ;
\end{tikzpicture}    
\caption{Typical finite volume in 2D}
\label{StencilFinitevolume}
\end{center}
\end{figure}

The basic governing equations considered for this study are compressible inviscid Euler equations. The conservative form of the equations are discretized using a finite volume method as in (\ref{EulerCompact}).  
\begin{subequations}\label{EulerCompact}
\begin{align}
\frac{d \overline{U}}{dt} = - R, 
R &= \frac{1}{\Omega} \left[\sum_{i=1}^{N}{F_c ~dS}\right], \\
 U =\left[\rho, \rho u,\rho v, \rho E_t\right]^T, F_c &=\left[\rho V_n, \rho u V_n + p \hat{n},  \rho v V_n + p \hat{n}, \left(\rho E_t + p \right) V_n \right]^T 
\end{align}
\end{subequations} 
where $U$ is conserved variable vector and $F_c$ is convective flux vector. The discretization is done using a 2-D finite volume method, as shown in figure (\ref{StencilFinitevolume}).  Here, $R$ represents the residue, denoting the flux gradient terms, $\Omega$ is the volume of cell with N faces, $V_n$ is the normal velocity $V_n = \Vec{V} \cdot \hat{n} =u n_x + v n_y$ on the control surfaces and $E_t = e + \frac{u^2 + v^2}{2}$ is the total energy per unit mass.  
These basic equations are not closed and require an EOS of the form $p = p\left(\rho, e\right)$ for closure. The simplest  EOS which represent a dense gas is van der Waals EOS given by \eqref{vweos}
\begin{align}
 \left( p + a \rho^2\right)\left(\frac{1}{\rho} - b \right) &= R T,\label{vweos} \\
 \textit{a} = 0.138 \times 10^{-3},~~
 \textit{b} &= 3.258 \times 10^{-5}.
\end{align} 
In (\ref{vweos}), the two terms $\textit{a}$ and $\textit{b}$ represent the pressure or intermolecular force and volume correction or co-volume parameter taking into consideration the intermolecular forces and the size of the molecules.  
\textcolor{black}{ The equation for internal energy takes the form \eqref{internale}
\begin{align}
\label{internale}
e = e_0 + C_v T - \alpha \rho
\end{align}
where, $e_0$ represent arbitrary reference value, $C_v$ is the specific heat at constant volume.
The non-dimensonalized  reduced variable form is obtained when the equation \eqref{vweos} is normalised with corresponding critical values $p_c,T_c~ \& ~\rho_c$ and a compressibility factor $Z_c = \frac{p}{\rho R T}=\frac{3}{8}$ resulting in 
\begin{align}
 \left( \bar{p} + a \bar{\rho}^2\right)\left(b-\bar{\rho} \right) &= 8 \bar{R} \bar{T},\label{rvweos} \\
 \textit{a} = 3,~~\text{b} &= 3.\\
 \bar{\delta} = \frac{\bar{R}}{\bar{C_v}}, \label{nddelta}
 \bar{e} = \frac{\bar{T}}{\bar{\delta}} -\frac{9}{8}\bar{\rho},\\
 \bar{a} = 1.5\sqrt{\frac{1}{\bar{\delta}}\left[ \left(1 + \bar{\delta} \right) \frac{4\bar{T}}{\left(3 - \bar{\rho}\right)^2} - \bar{\rho} \right] },\label{ndsos}\\
\Gamma = \frac{6}{\bar{\delta}\bar{a}^2}\left[\left(\frac{1}{3 - \bar{\rho}}\right)^3 A  \bar{T} - \frac{\bar{\rho}}{4}\right], A = 2 + 3\bar{\delta} + \bar{\delta}^2 ,\label{ndfd}
\end{align} 
}
\subsection{Discretization strategies}  
The governing equations are solved using a cell centered finite volume method  with cell integral averages defined as in \eqref{averageU}. For the sake of simplicity, the discretization strategies are explained for 1D case in the following.  The finite volume update formula for Euler equations is given by \eqref{1DEulerupdate} with interface flux evaluated as in \eqref{interfaceflux1} 
\begin{align}
 \overline{U} &= \frac{1}{\Omega}\int_{\Omega} U d\Omega~.\label{averageU}\\
 \bar{U}^{n+1}_j &=    \bar{U}^{n}_j - \frac{\Delta t}{\Delta x}\left[ F^n_{j + \frac{1}{2}} - F^n_{j - \frac{1}{2}} \right] \label{1DEulerupdate}\\
\label{interfaceflux1}
 F_{j \pm \frac{1}{2}} = F_I\left(U_L,U_R \right) &= \frac{1}{2}\left[ F(U_L) + F(U_R) \right] - \Delta F_{num} 
\end{align}
where the first term on the right hand side is an average flux from the left (L) and the right (R) states and $\Delta F_{num}$ is a flux difference  representing numerical diffusion. The numerical diffusion flux can be written as in \eqref{DiffusionFlux}   
\begin{equation} \label{DiffusionFlux}
\Delta F_{num} =d_I = \frac{\mid \alpha_{\mathrm{I}} \mid}{2}\left(U_R - U_L\right) 
\end{equation}
where $\alpha_{\mathrm{I}}$ coefficient of numerical diffusion $d_I$.  Most of the numerical methods differ in the way this coefficient is determined. In the present work, $\alpha_{\mathrm{I}}$, is obtained by algorithms RICCA and MOVERS+ \cite{Ramesh_FP},\cite{rkolluru}. The details of these algorithms are explained in the following sections.
\subsection{Why do we need algorithm independent of EOS?} 
In this section we present the analysis on why an algorithm is required to be independent of EOS.  Let us consider the governing equations in 1D with general EOS of the form $p = p \left(\rho,e\right)$.  In nonconservative form these equations can be written as in \eqref{1Dnonconservative} where A(U) is the flux Jacobian matrix \eqref{1Dfluxjacobian}.
\begin{align}
 \label{1Dnonconservative}
 \frac{\partial U}{\partial t} + A\left(U\right)\frac{\partial U}{\partial x} &= 0\\
 \label{1Dfluxjacobian}
 A\left( U \right) &= \frac{\partial F}{\partial U}
\end{align}
For 1D Euler equations flux Jacobian matrix is given by \eqref{1dmatrixfluxjacobian} with total enthalpy $H = E_t + \frac{p}{\rho}$
\begin{align}
\small
\label{1dmatrixfluxjacobian}
 A(U) = \frac{\partial F(U)}{\partial U}= \left[\begin{array}{ccc}
0 & 1 & 0 \\
\left(a^2 - u^2 \right) - \frac{1}{\rho}\frac{\partial p}{\partial e}\left(H -u^2\right) & 2 u - \frac{u}{\rho} \frac{\partial p}{\partial e} & \frac{1}{\rho} \frac{\partial 
p}{\partial e}  \\
\left(a^2 - H \right)u - \frac{u}{\rho}\frac{\partial p}{\partial e}\left(H -u^2\right) & H - \frac{u^2}{\rho} \frac{\partial p}{\partial e} & u + \frac{u}{\rho} \frac{\partial 
p}{\partial e}  \\
\end{array}\right] 
\end{align}
It can be observed that the flux Jacobian matrix with general EOS is a function of  $\frac{\partial p}{\partial e}$. The eigenvalues for the flux 
Jacobian matrix are 
\begin{align}
 \lambda_1 = u -a ;~~ \lambda_2 = u ;~~ \lambda_3 = u + a
 \end{align}
 with the corresponding eigenvectors being
 \begin{align}
  R^1 = \left[ \begin{array}{c} 1\\  u-a  \\ H -ua \end{array}\right] ; R^2= \left[ \begin{array}{c} 1\\ u \\ H -\frac{\rho a^2}{\left(\frac{\partial p}{\partial e}\right)}  \end{array} 
\right] ; R^3 = \left[ \begin{array}{c} 1\\  u+a  \\ H + ua \end{array}\right] ;
 \end{align}
The acoustic speed or sound speed, $a$, with general EOS is given by \eqref{speedofsound} or alternatively by \eqref{alternatespeedofsound}
\begin{align}
\label{speedofsound}
a &= \sqrt{ \frac{p}{\rho^2} \frac{\partial p}{\partial e} + \frac{\partial p}{\partial \rho}}\\
a &= -\frac{C_p}{C_v} \frac{\frac{\partial T}{\partial \rho}}{\frac{\partial T}{\partial p}}
\label{alternatespeedofsound} 
\end{align}
 It can be observed that, speed of sound $a$ as given in \eqref{speedofsound}, is a function of the derivatives $\left(\frac{\partial p}{\partial e}~ \&~ \frac{\partial p}{\partial \rho}\right)$ and is strongly related to the formulation of EOS. If an analytic expression for EOS exists then the derivatives can be found and hence the eigen-structure of the hyperbolic system can be derived. \textcolor{black}{If the EOS is complicated or if the estimation of the derivatives is not possible analytically, then estimating the eigen-structure of the flux Jacobian matrix might be complicated or may not even be possible.} Hence, if the numerical schemes depend upon the eigen-structure of the flux Jacobian matrix, for example as in Riemann solvers and flux vector splitting methods, then the development of the numerical method for real gases becomes complicated or some times even impossible. Most of the upwind schemes like those of Steger-Warming and van Leer require the complete knowledge of eigen-structure of the flux Jacobian matrix, evaluation of which might become complicated based on the nature of EOS. Thus substantial modifications are to be carried out for the upwind schemes when applied to real gases. 

The numerical schemes \textcolor{black}{described} in the following sections, MOVERS+ and RICCA, do not depend strongly on the eigen-structure of the underlying hyperbolic system, especially on the eigenvectors. \textcolor{black}{For these numerical schemes, except for evaluation of speed of sound, the knowledge of the eigenvalues and eigenvectors is not required. Thus application of these central solvers to any hyperbolic system with real gas EOS is straightforward and requires no modification at all \cite{rkolluru}.}

\section{Riemann Invariants based Contact-discontinuity Capturing Algorithm (RICCA)}\label{sec:RICCA}
In this section a novel scheme developed in \cite{Ramesh_FP},\cite{rkolluru} is described, which is based on Generalized Riemann Invariants (GRI). When the concept of GRI is utilized in the discretization process it leads to a scheme which captures steady contact discontinuities exactly. 

\begin{figure}[h!]
\begin{center}
\begin{tikzpicture}[scale = 1.3]
\draw (2,1)node (xaxis) [below] {(j)} -- (4,2)  node (xaxis) [below] {(j+1)};
\draw [->,red,very thick](3,1.5) -- (3.4,1.7) node (xaxis) [above] {$\hat{n} $};
\draw [red](2,0.5) node (xaxis) [below] {$L$}; 
\draw [red](4,2.5) node (xaxis) [below] {$R$};
\draw (2,1.1) node (xaxis)[left] {$F_L = F(U_L)$}; 
\draw (4,2.1) node (xaxis)[right] {$F_R = F(U_R)$};
\draw [red,thick,dashed](4,1) node (yaxis) [below] {$j+\frac{1}{2}$}-- (2,2);
\draw[red](2,2) node (yaxis) [above] {$F_\mathrm{I}$};
\draw [blue,fill] ( 3,1.5) circle [radius=0.03] ;
\draw [blue,fill] ( 2,1) circle [radius=0.03] ;
\draw [blue,fill] ( 4,2) circle [radius=0.03] ;
\end{tikzpicture}    
\caption{Typical cell-interface of a 1D finite volume}
\label{Interfaceflux1}
\end{center}
\end{figure}
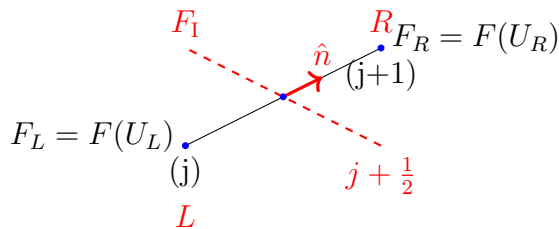

 The modelling of the numerical diffusion is carried out as follows. Consider an interface of a control volume, as shown in figure (\ref{Interfaceflux1}), across which the flow is assumed to be 1D and diffusion flux is to be evaluated. Evaluating the flux difference is as given in \eqref{flxudiff}.
\begin{equation} 
\label{flxudiff} 
\Delta F_{num} = \left( \displaystyle \frac{\Delta F}{\Delta U} \right)_{num} \Delta U = \alpha_{num} \Delta U = \alpha_{I} \Delta U 
\end{equation} 
The coefficient of numerical diffusion, $\alpha_{num}$, is modelled using a diagonal matrix as explained in \cite{Jaisankar_SVRRao}.
	  Various numerical schemes differ in the way the wave speed or the coefficient of numerical diffusion is determined. The basic idea of the present work is to use Generalized Riemann Invariant (GRI) across the interface to determine the coefficient of diffusion, $\alpha_{I}$.  When a GRI is applied to a contact discontinuity we obtain \eqref{diffusionu}

\begin{equation}\label{diffusionu}
\alpha_1 = \alpha_2 = \alpha_3 =\alpha_I = |u|
\end{equation}	   
  which for any arbitrary interface as given in figure (\ref{Interfaceflux1}) can be given by \eqref{diffusionVn}
  \begin{equation}\label{diffusionVn}
\alpha_I = |V_n| = |V_{n_L}| = |V_{n_R}| = \frac{|V_{n_L}| + |V_{n_R}|}{2} = max( |V_{n_L}|, |V_{n_R}|)
\end{equation}	
  
 Numerical experimentation has revealed that, this numerical diffusion evaluated by \eqref{diffusionVn}, though adequate in capturing the contact-discontinuities exactly, is not sufficient enough for the case of shocks being located at the cell interface. So in order to generalize the diffusion for any case, the Riemann Invariants based Contact-discontinuity Capturing Algorithm (RICCA) is designed with the following coefficient of numerical diffusion:
\begin{equation}\label{accudisks_eval_alpha_euler_2d}
{\alpha}_ {\emph{$I$}} = \begin{cases}
                            \qquad \qquad \frac{|V_{n_L}|+|V_{n_R}|}{2}, \qquad \qquad \qquad \quad \text{if } |\Delta\mathbf{F}|<\epsilon \ \text{and} \ |\Delta\mathbf{U}| <\epsilon, \epsilon = 1\times 10^{-8}\\
                            \quad \qquad \qquad \qquad \qquad \qquad \qquad \qquad \qquad  \\
                            max(|V_{n_L}|, |V_{n_R}|) + sign(|\Delta p_{\raisebox{-2pt} {\scriptsize {$\mathrm{I}$}}}|) a_{\mathrm{I}} , \qquad \text{otherwise} \\
                            \qquad \qquad \qquad \qquad \quad
                           \end{cases}
\end{equation}
where $\epsilon$ is a small number and  $a_{\mathrm{I}}$ is the speed of sound evaluated based on the EOS under consideration  given by
\begin{align}
a_I &= \frac{a_L+a_R}{2}, \\
\Delta{p}_{\raisebox{-2pt} {\scriptsize \emph{$I$}}} &= (p_R-p_L).
\end{align}

This design of the coefficient of numerical diffusion therefore does not require any entropy fix.

Features of the new central scheme RICCA are as follows. It can capture steady grid-aligned contact-discontinuities exactly.  It has sufficient numerical diffusion near shocks so as to avoid shock instabilities.  It does not need entropy fix at sonic points.  It is not tied down to the eigen-structure and hence can be easily extended to any general equation of state, without modification. 

\section{Central solver \emph{MOVERS+} }\label{sec:movers+}
The second of the algorithm utilized in this work is based on modification of a central Rankine-Hugoniot solver as in \cite{Jaisankar_SVRRao}, called as MOVERS (Method of Optimal Viscosity for Enhanced Resolution of Shocks). 
MOVERS  requires wave speed correction in order to restrict the coefficient of diffusion to within the eigenspectrum. To avoid wave speed correction, a simpler strategy is proposed in this section which is described below.
\begin{align}
\label{alphai1}
d_{\mathrm{I,j}} &=  \frac{1}{2}\left| \frac{\Delta F_j}{\Delta U_j}\right| \Delta U_j, \quad j =1,2,3 \\
	 &= \frac{1}{2}\frac{\left |\Delta F_j \right |}{Sign(\Delta U_j) \Delta U_j } \Delta U_j  \\ 
	 & = \frac{1}{2} Sign\left(\Delta U_j\right) \left| \Delta F_j \right| , \quad j =1,2,3
\end{align} 
where the relation $\frac{1}{Sign(\cdot)} = Sign(\cdot)$ is used.  
This form of $d_I$ will eliminate the need of wave speed correction for MOVERS. Numerical experimentation has revealed that this numerical scheme has very low diffusion and captures steady discontinuities exactly but encounters problems in smooth regions due to lack of sufficient numerical diffusion. Therefore, using a shock sensor (\ref{ShockSensor}), an additional numerical diffusion is introduced. This additional diffusion is based on the fluid velocity, which is demonstrated to be sufficient to avoid unphysical expansions in smooth regions \cite{Ramesh_FP,rkolluru}.  
The coefficient of numerical diffusion for MOVERS+ is given by
\begin{align}
\label{MOVERS_NWSC}
  d_{\mathrm{I}}  =  \frac{\Phi}{2} Sign(\Delta U_j)\lvert \Delta F_j \rvert + \left(\frac{|V_{nL}|+|V_{nR}|}{2}\right) \Delta U_j, \quad j =1,2,3 
\end{align} 
where the $\Phi$ is the shock sensor defined by  
\begin{align}
\label{ShockSensor}
\Phi = \left| \frac{\Delta p}{2 {p_{\mathrm{I}}}}\right| \ \textrm{with} \ p_{I} = \frac{p_L+p_R}{2} 
\end{align}

The features of this modified algorithm, MOVERS+, are as follows.  It can capture steady grid-aligned contact discontinuities exactly and provides low diffusion otherwise.  It has sufficient numerical diffusion near shocks so as to avoid shock instabilities (deliberately giving up exact shock capturing for gain in robustness).  It does not need entropy fix for smooth regions or in expansion regions.  It does not require any wave speed correction. It is a simple central solver and is not based on Riemann solvers, field-by-field decompositions or  complicated flux splittings, thus making it a suitable candidate for further extensions.  

It can be observed that the above numerical schemes do not depend on the EOS and on the eigen-structure of the underlying hyperbolic systems. If the information of the properties in the cell or the control volume are known then they can be utilized directly. Hence these algorithms can be extended to any form of EOS.

\section{Test cases and results for dense gas} \label{sec:results}
To test the capabilities of the algorithms mentioned in the previous sections some interesting 1D and 2D test cases are carefully chosen where the physics of the problem is utmost important and the non-classical behaviour of the gases is clearly predominant. Numerical simulations are carried out with MOVERS+ and RICCA in 1D and 2D results of MOVERS+ and RICCA are presented in this section.
\subsection{1D dense gas test cases}
In this section numerical simulations for 1D dense gases are presented. These test cases are taken from \cite{argrow,cinnella} and use van der Waals EOS for simulations. Three test cases are considered to study the non-classical wave phenomena and are chosen such that the flow field contains some regions of negative fundamental derivative ($\Gamma < 0 $). For all the three test cases the number of cells used for computational study are $n = 500$ with $CFL = 1$, and the solutions are presented at the times given in table (\ref{dgtable:1}).
  Initial conditions used for shock tube test case are described in table (\ref{dgtable:1}) and results are compared with the reference data obtained from \cite{argrow}.

\begin{figure}[htbp!]
\begin{center}
  \includegraphics[scale = 0.6,keepaspectratio]{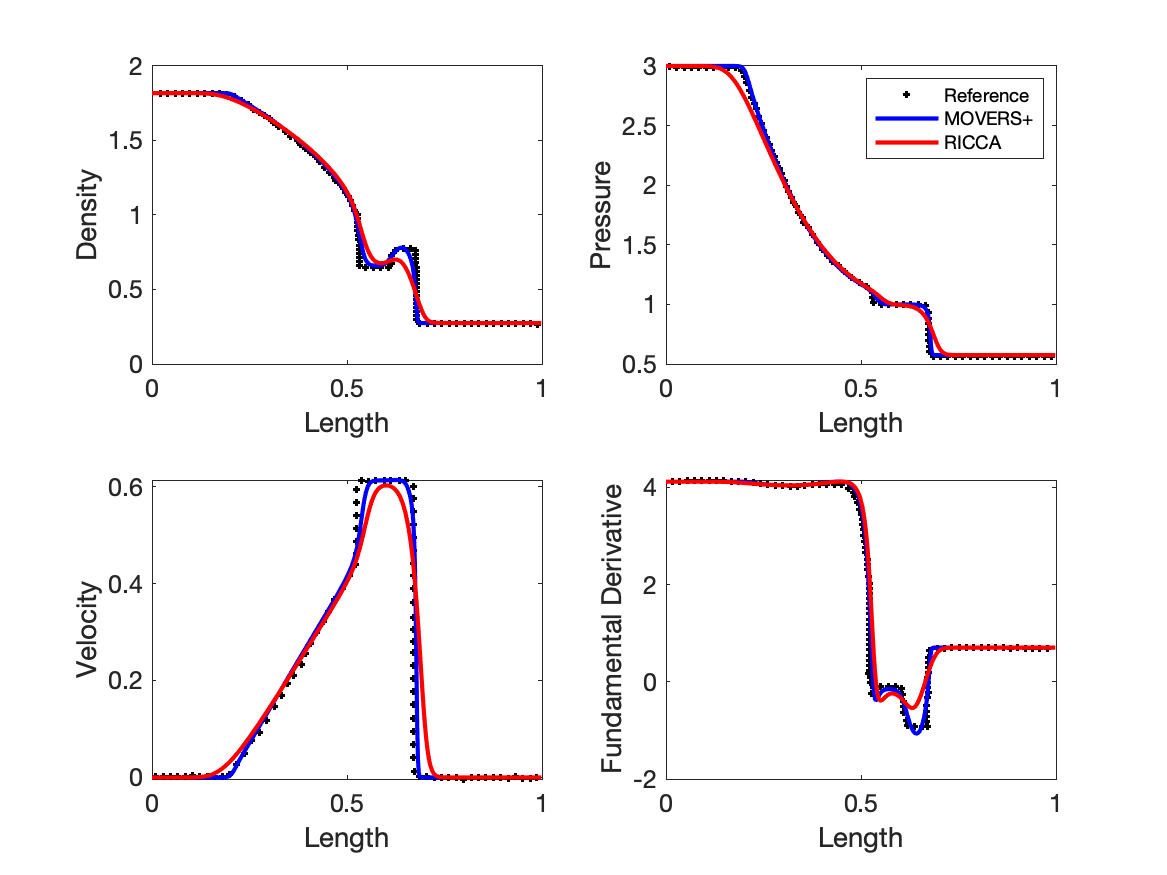} 
  \caption{Dense gas test case 1 with MOVERS+ and RICCA with n = 500 at t=0.1807}
  \label{dg1_all}
\end{center}
\end{figure}

\begin{table}[htbp!]
\centering
\begin{tabular}{ |c|c|c|c|c|c|c|}
\hline
Test case     &    $\delta$         &         $\rho_L$         &     $p_L$     &     $\rho_R    $     &     $p_R$     &     Time \\
\hline
1        &    0.0125    &    1.818    &    3.0        &    0.275    &    0.575    &    0.1807\\
\hline
2        &    0.0125    &    0.879    &    1.090    &    0.562    &    0.885    &    0.4801\\
\hline
3        &    0.0125    &    0.879    &    1.090    &    3.630    &    0.575    &    0.2917\\
\hline
\end{tabular}
\caption{Test cases for simulation of dense gas flows}
\label{dgtable:1}
\end{table}
%
%
%
%
%
%
Test case 1 in table (\ref{dgtable:1}) represents a Riemann problem where initially both the left and the right states lie within $(\Gamma > 0)$ region. The solution has a left moving rarefaction fan, a contact-discontinuity in the middle and a right moving shock wave. Initially the rarefaction fan is in $(\Gamma > 0)$ region. However, as the flow evolves $\Gamma$ changes its sign to $(\Gamma <0)$ and rarefaction fan steepens into a rarefaction shock.  Figure (\ref{dg1_all}) refers to the numerical simulation of this test case using RICCA, and MOVERS+.  It can be observed that all the numerical methods are resolving the fundamental features accurately. It can also be seen that though RICCA is slightly diffusive in nature, can be utilized in predicting the flow features well. Further no oscillations are observed from both MOVERS+ and RICCA.

\begin{figure}[htbp!]
\begin{center}
    \includegraphics[scale=0.6,keepaspectratio]{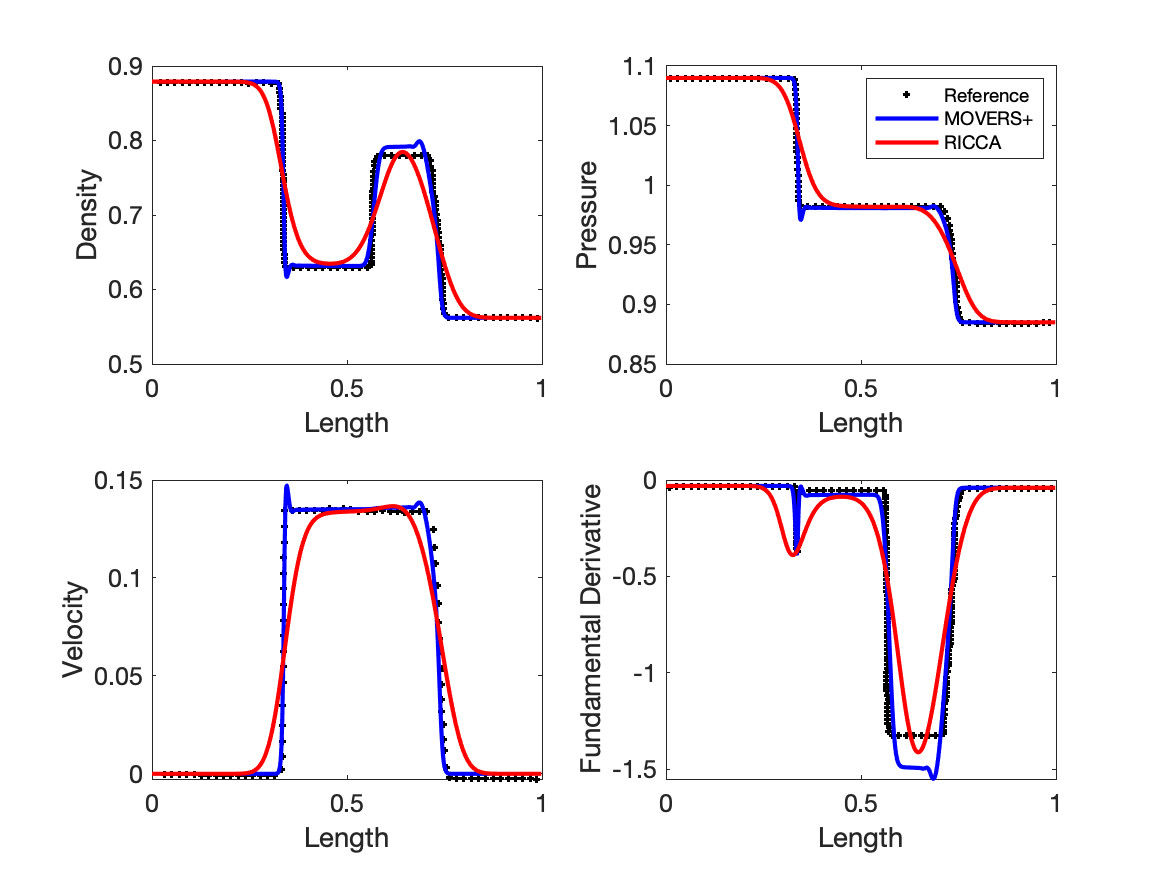} 
\caption{Dense gas test case 2 with MOVERS+ and RICCA with n = 500 at t = 0.4801}    
\label{dg2_all}
\end{center}
\end{figure}

Test case 2 in the table (\ref{dgtable:1}) has both left and right states lying in the $\Gamma <0$ region. The fundamental derivative remains negative everywhere and the flow behaviour is exactly opposite with respect to classical Riemann problem. Specifically, the solution presents a left running rarefaction shock, a middle contact discontinuity, and a right running compression fan. Except for the fact that $\Gamma$ is negative in the entire domain for this test case, the non-linear waves are similar to those in the classical Riemann problem. Figure (\ref{dg2_all}) refers to the numerical simulation of this test case using RICCA,  and MOVERS+.

\begin{figure}[htbp!]
\begin{center}
  \includegraphics[scale=0.6,keepaspectratio]{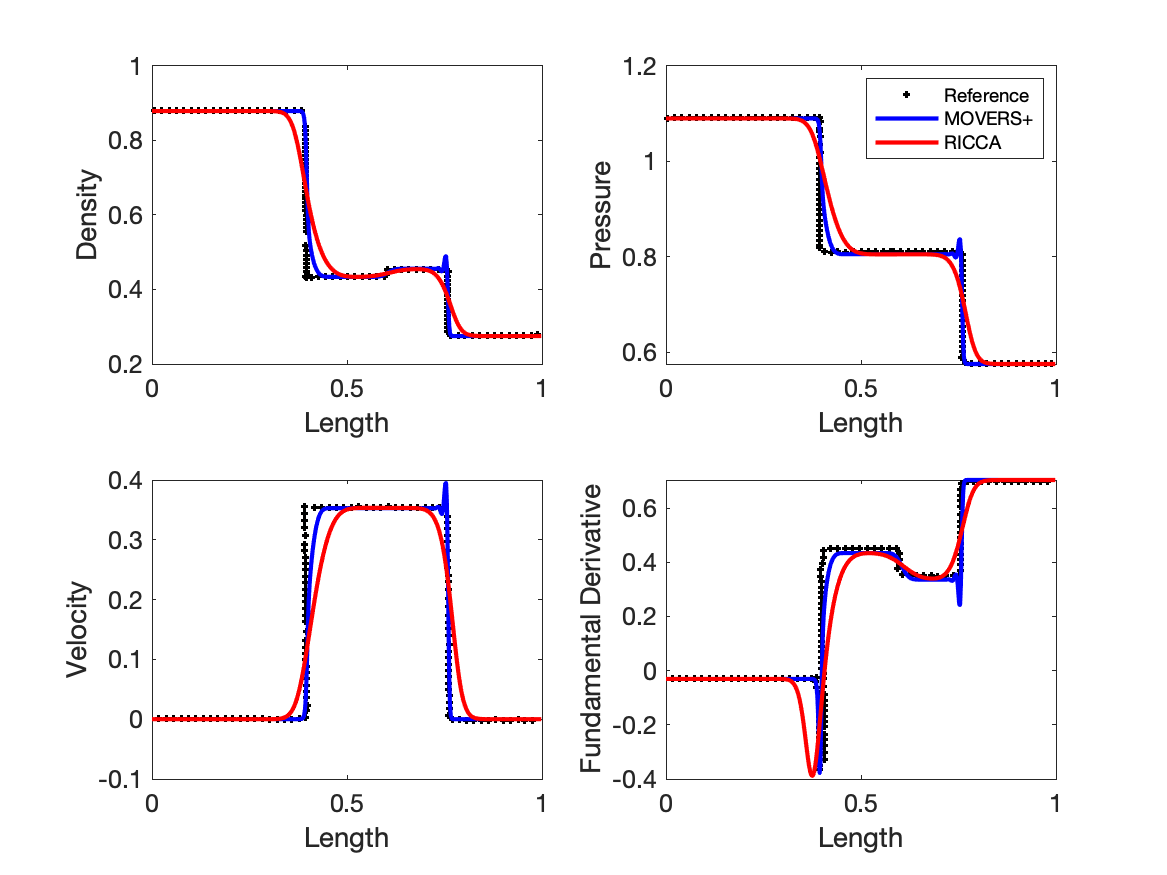} 
  \caption{Dense gas test case 3 with MOVERS+ and RICCA with n = 500 at t = 0.2917}    
  \label{dg3all}
\end{center}
\end{figure}

Test case 3 is an example of flow evolution with two states corresponding to $(\Gamma <0)$ and $(\Gamma > 0)$. The fundamental derivative changes its sign from left to right and a mixed rarefaction wave forms at the crossing of the transition line. The compression wave lies entirely within the classical zone and hence a classical shock wave appears in this region. Figure (\ref{dg3all}) refers to the numerical simulation of test case 3 using RICCA and MOVERS+. 
From the above three test cases it can be inferred that all the numerical algorithms are capable of capturing the non-classical wave phenomena with reasonable accuracy without any modification to the basic algorithm. 

\subsection{2D dense gas test cases and results}
To analyse the non-classical behaviour of dense gases in 2D both steady state and transient test cases on simple geometries are considered. To distinguish between classical and non-classical waves both perfect gas EOS and van der Waals EOS are used for comparison. Algorithms RICCA and MOVERS+ are used in these simulations and their ability to capture the non-classical phenomena is thoroughly explored.  Only second order accurate solutions are presented.

 \subsection{Steady state test cases}
The following steady state test cases are considered.
\begin{itemize}
\item Supersonic flow over a forward facing step
\item Supersonic flow over a circular arc.
\item Supersonic flow over a smoothly varying expansion ramp
\end{itemize}
Initial conditions for steady state cases in table (\ref{dgtable:2DSteadystatecases}). Steady state conditions for the dense gases are represented as DGS1, DGS2 and DGS3.  
\begin{table}[htbp]
\centering
\begin{tabular}{ |c|c|c|c|c|c|c|}
\hline
test case     &    $\delta$         &         $\rho$         &     $p$     \\
\hline
DGS1        &    0.0125    &        1.00            &    1.00    \\
\hline
DGS2        &    0.0125    &        0.88            &     1.09\\
\hline
DGS3        &    0.0125    &        0.62            &     0.98\\
\hline
\end{tabular}
\caption{Free stream conditions for steady state test cases for simulation of dense gas flows}
\label{dgtable:2DSteadystatecases}
\end{table}
\textcolor{black}{For perfect gas EOS the value of $\gamma$ is taken as $1.4$ and for van der Waals gas the value of $\delta = \frac{R}{C_v} = 0.0125$ is maintained constant which corresponds to value of $\gamma = 1.1025$ .}

\subsubsection{Supersonic flow over a forward facing step}
\begin{figure}[!ht]
\makebox[\textwidth][c]{%
\begin{subfigure}[b]{0.4\textwidth}
\centering
\includegraphics[scale=0.1,keepaspectratio]{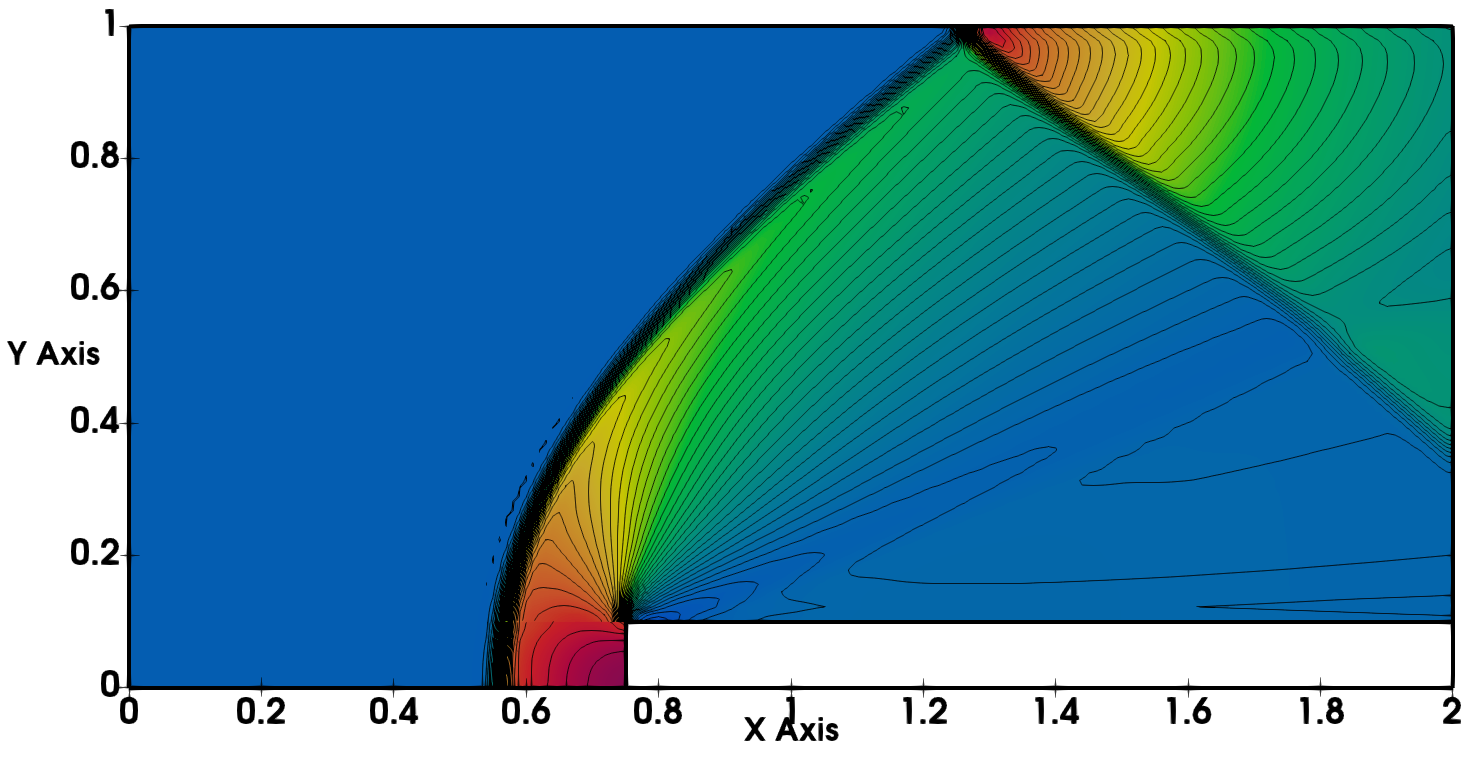}
\caption{Perfect Gas}
\label{FS_D_PGEOS}
\end{subfigure}%
\begin{subfigure}[b]{0.4\textwidth}
\centering
\includegraphics[scale=0.14,keepaspectratio]{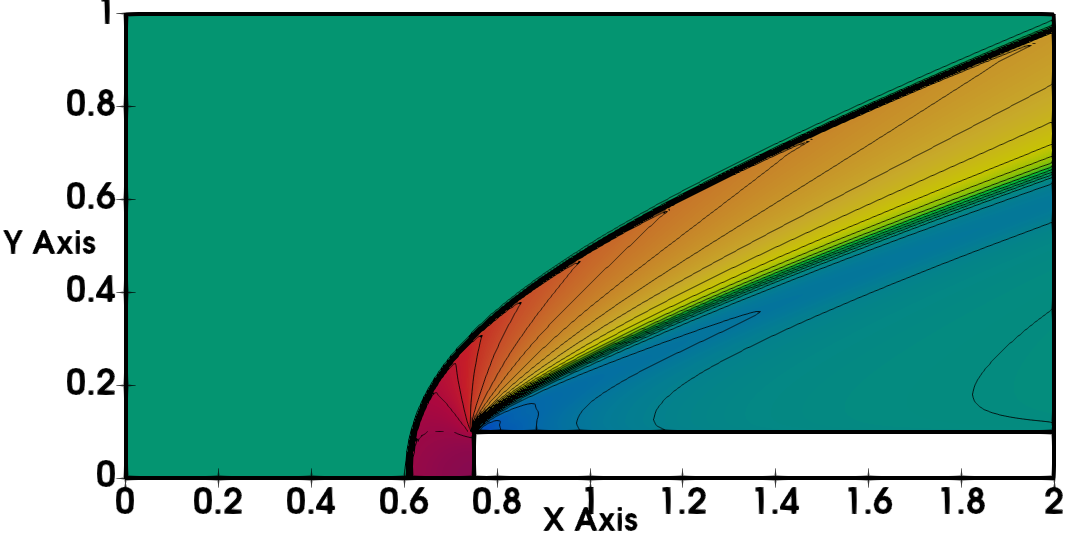}
\caption{van Der Waals EOS }
\label{FS_D_VWaal}
\end{subfigure}%
\begin{subfigure}[b]{0.4\textwidth}
\centering
\includegraphics[scale=0.1,keepaspectratio]{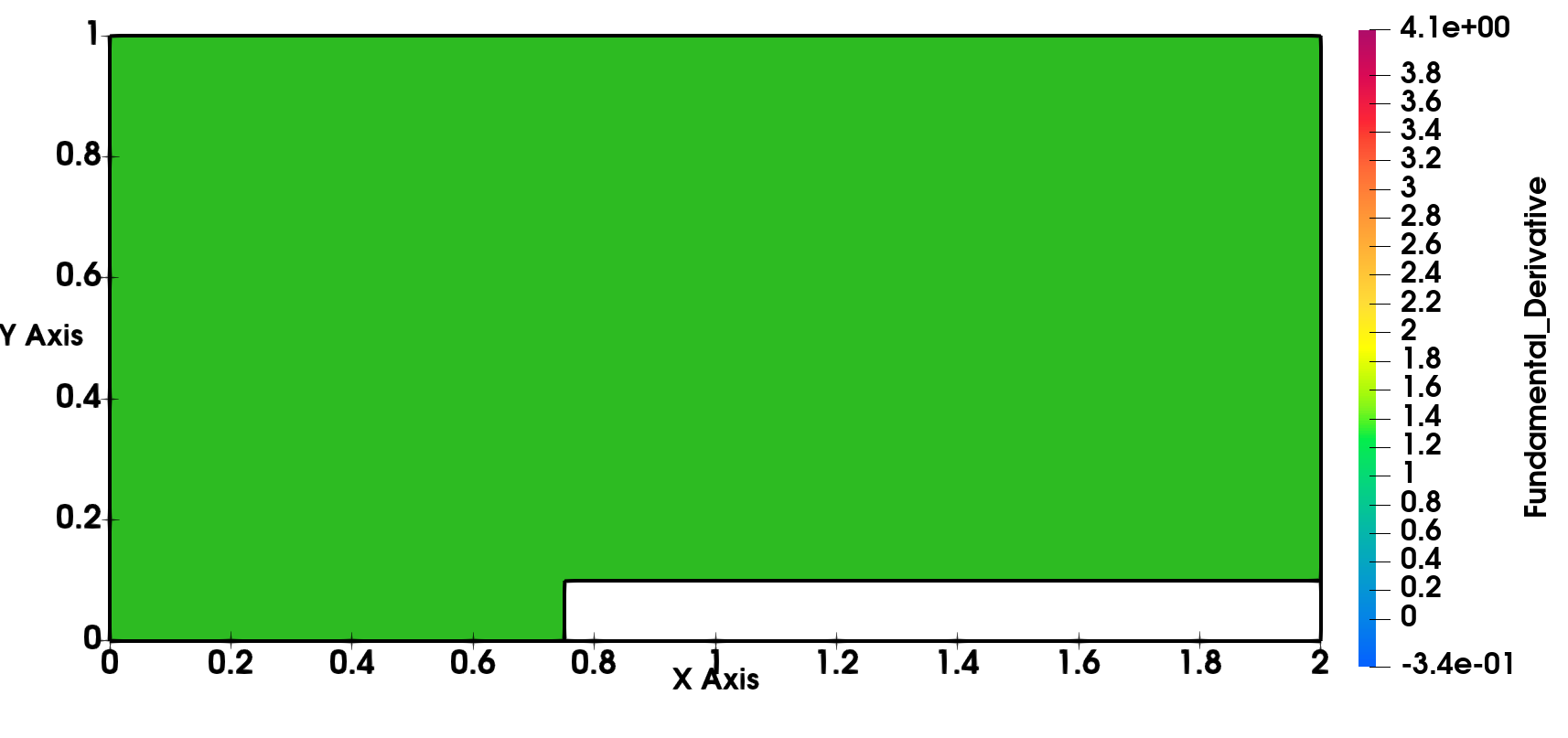}
\caption{Fundamental Derivative}
\label{FS_FD_PGEOS}
\end{subfigure}

}
\caption{$M=3.0$ flow on forward step with perfect gas EOS and van Der Waals EOS}
\label{FS_PGEOS}
\end{figure}

A steady supersonic flow of perfect gas and dense gas over a forward facing step is considered here. The computational domain consists of $[0,2]\times[0,1]$ with a step height of 0.2m placed at $x=0.75$. The left boundary is considered as supersonic inlet, right boundary as supersonic exit, top and bottom boundaries as inviscid walls. 
 For perfect gas the incoming supersonic flow $(M = 3.0)$ encounters a forward facing step. A detached bow shock is formed because of the step and terminates into an oblique shock reflection on the upper boundary. The reflected oblique shock exits out of the boundary as shown in Figure (\ref{FS_PGEOS}).  An expansion fan centred at the corner of the step evolves and interacts with the reflected shock. It can be observed that the fundamental derivative does not change its sign as evident from Figure (\ref{FS_FD_PGEOS}). \textcolor{black}{Figure(\ref{FS_D_VWaal}) shows the density contours of dense gas with $M=3.0$ on a forward step. It can be observed from the figure that the flow structure is significantly different from that of the perfect gas EOS. It can also be observed that the shock doesn't hit the upper wall at all though instead of an expansion fan an expansion shock is observed. }

\begin{figure}[h!]
\makebox[\textwidth][c]{%
\begin{subfigure}[b]{0.4\textwidth}
\centering
 \includegraphics[scale=0.1,keepaspectratio]{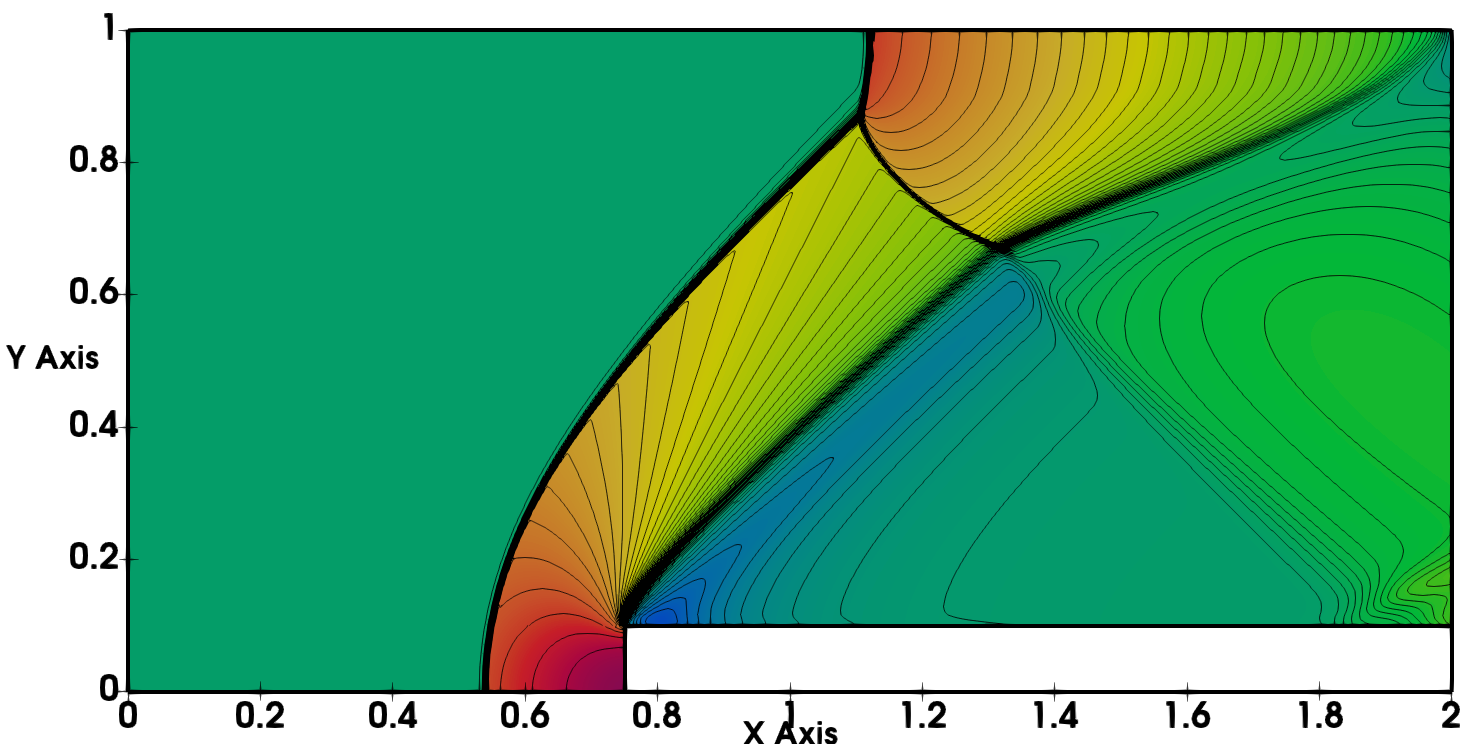}
\caption{Density contours}
 \label{FS_DC_VWEOS_1000x500}
\end{subfigure}%
\begin{subfigure}[b]{0.4\textwidth}
\centering
    \includegraphics[scale=0.1,keepaspectratio]{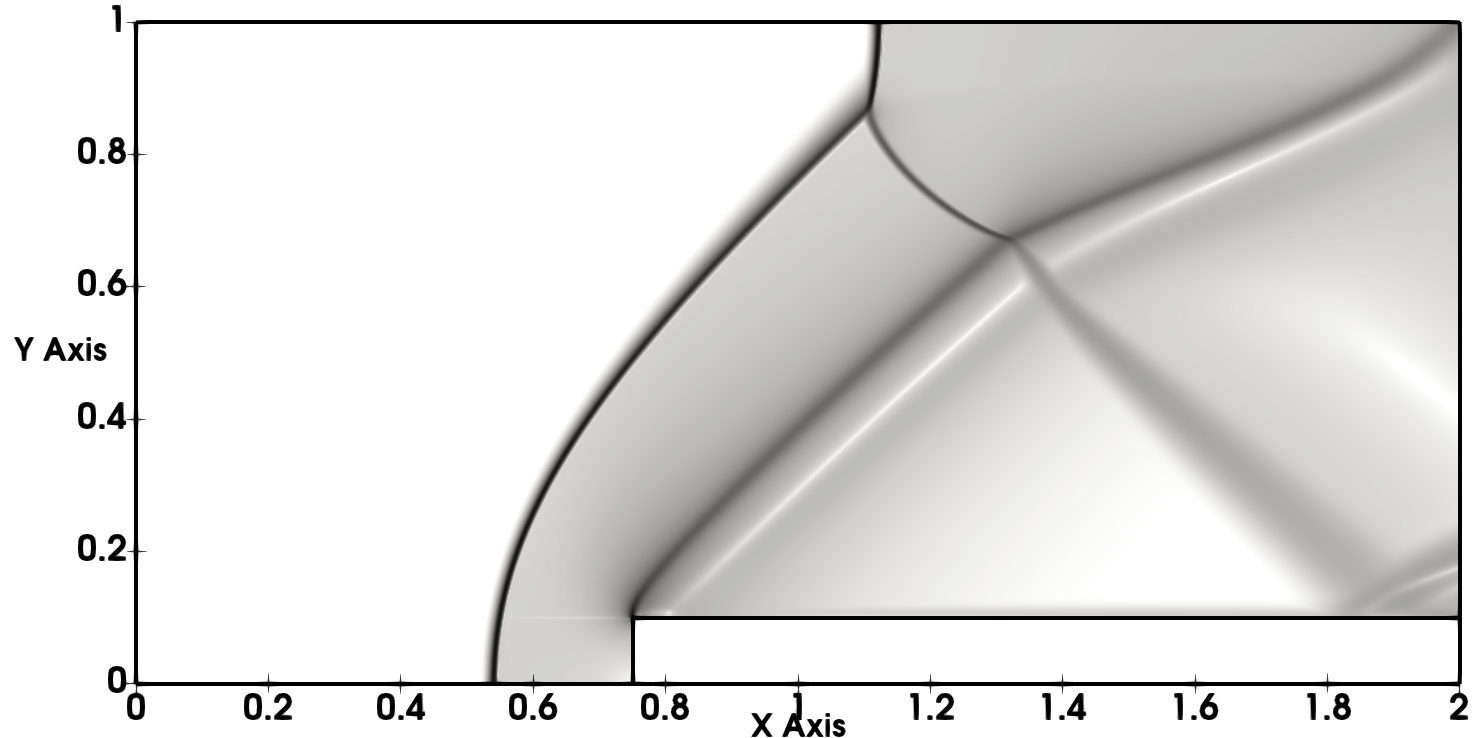}
\caption{Density Gradients}
\end{subfigure} 
 \begin{subfigure}[b]{0.4\textwidth}
\centering
    \includegraphics[scale=0.1,keepaspectratio]{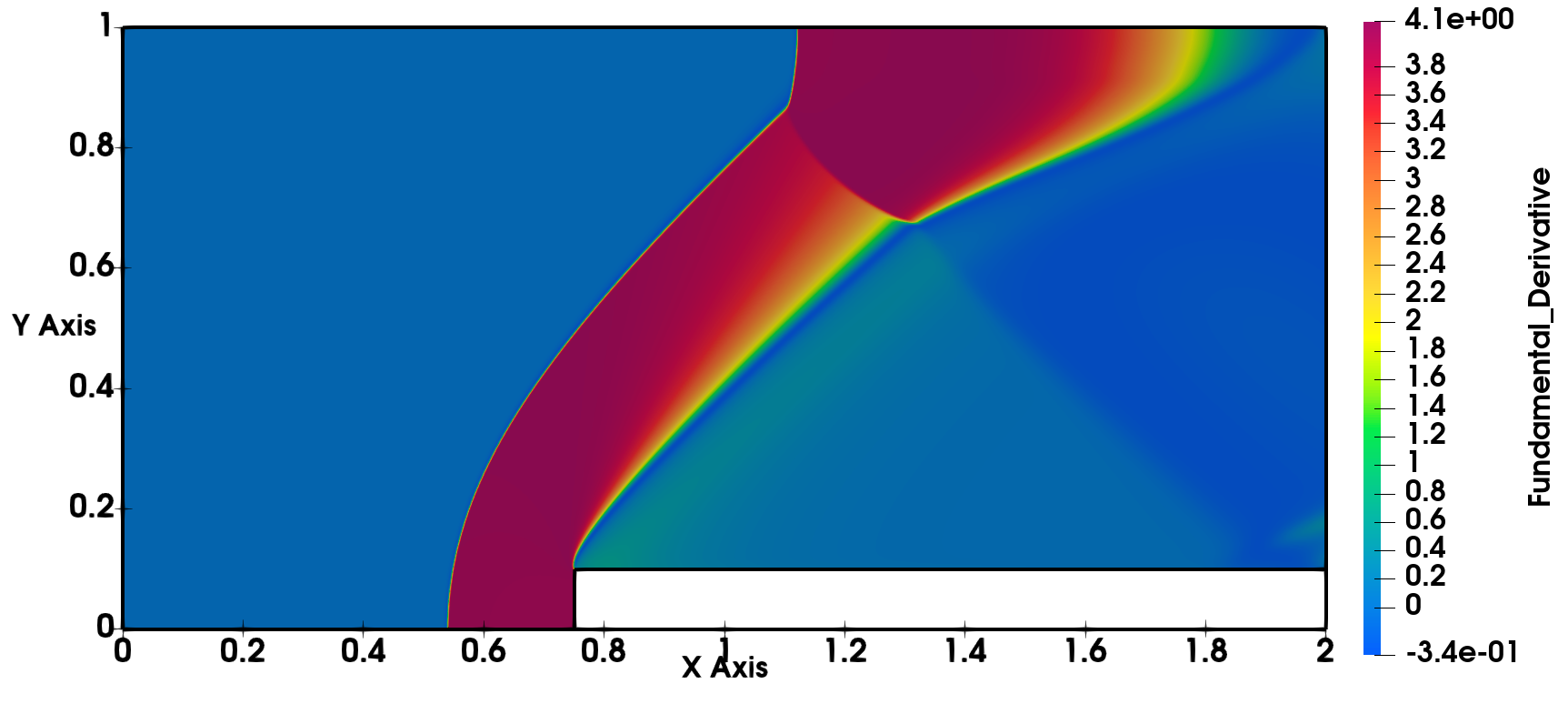}
\caption{Fundamental Derivative}
 \label{FS_FD_VWEOS_1000x500}
\end{subfigure}

 }

  \caption{$M=1.5$ flow on Forward facing Step using van Der Waals EOS and MOVERS+ on $1000\times500$ grid}
   \label{FS_VWEOS_1000x500}
\end{figure}

\begin{figure}[!htb]
\makebox[\textwidth][c]{%
\begin{subfigure}[b]{.4\textwidth}
\centering
 \includegraphics[scale=0.1,keepaspectratio]{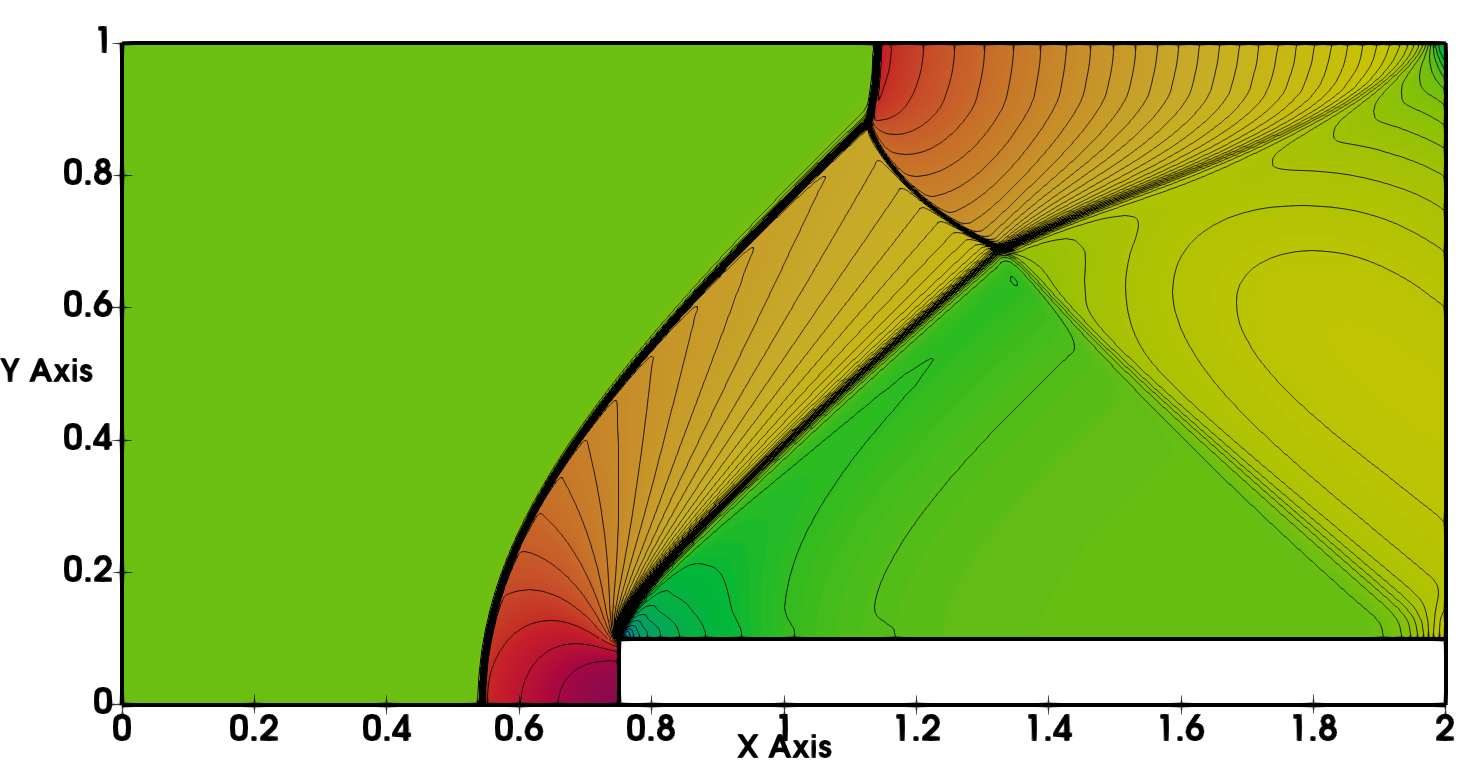}
\caption{Density contours}
 \label{FS_DC_VWEOS_1000x500_RICCA}
\end{subfigure}%
\begin{subfigure}[b]{.4\textwidth}
\centering
    \includegraphics[scale=0.1,keepaspectratio]{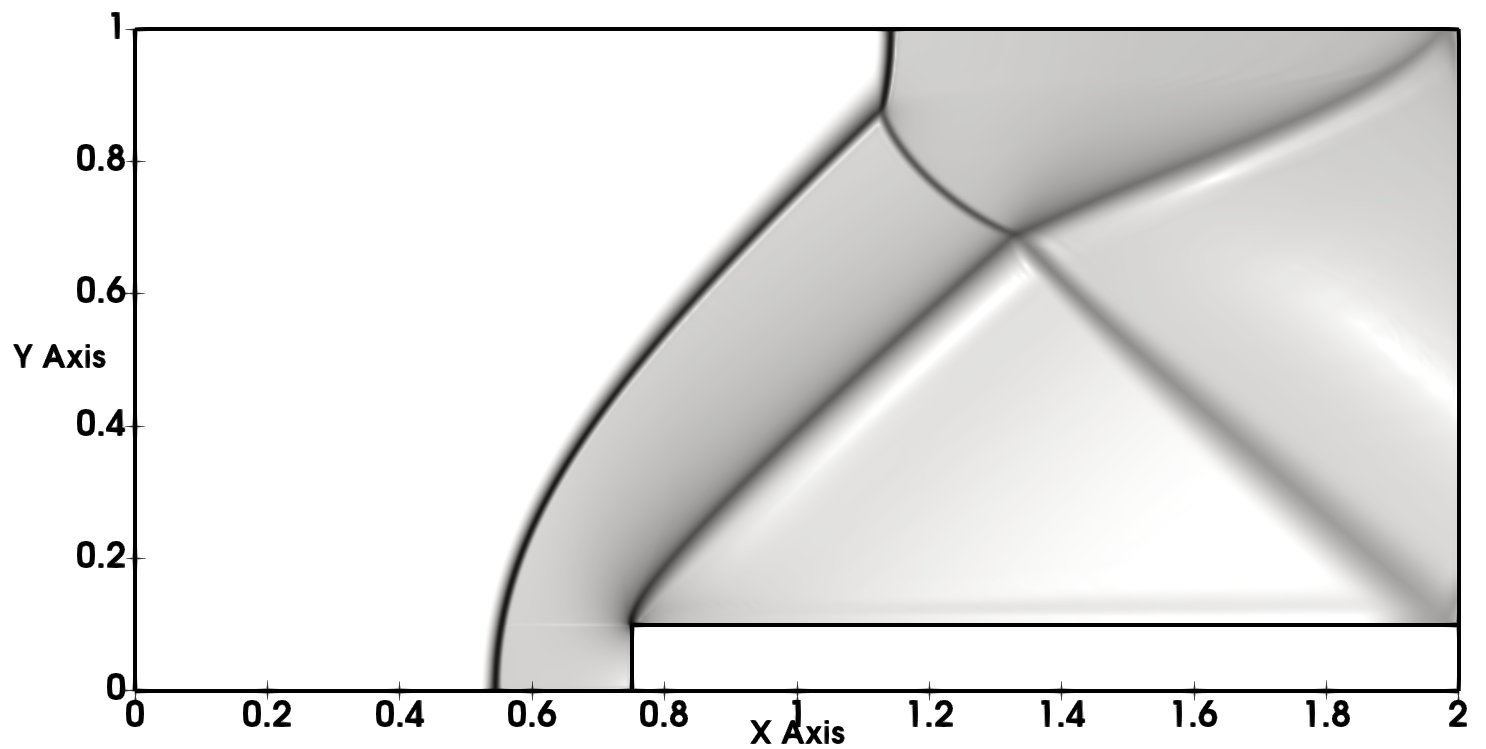}
\caption{Density Gradients}
\end{subfigure} 
\begin{subfigure}[b]{0.4\textwidth}
\centering
    \includegraphics[scale=0.1,keepaspectratio]{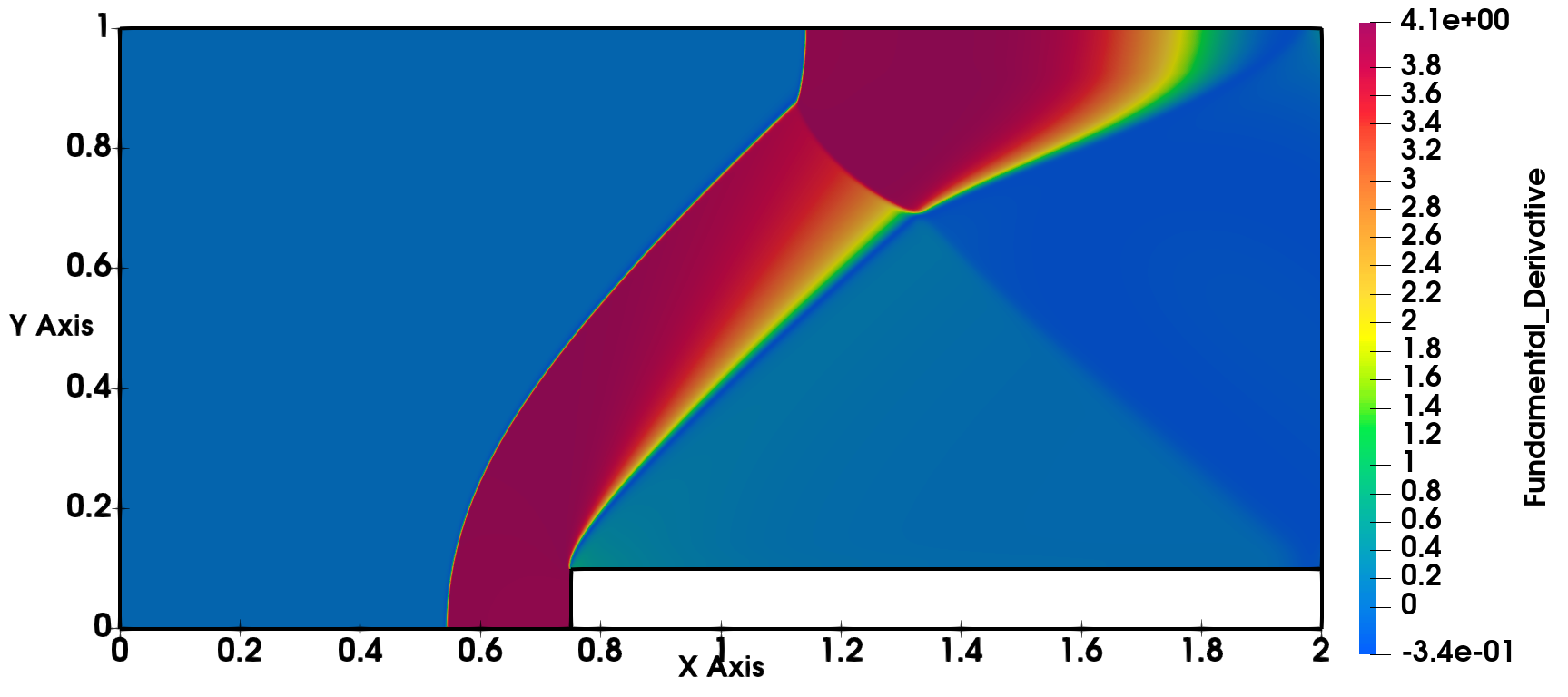}
\caption{Fundamental Derivative}
 \label{FS_FD_VWEOS_1000x500_RICCA}
\end{subfigure}
 }

  \caption{$M=1.5$ flow on Forward facing Step using van Der Waals EOS and RICCA on $1000\times500$ grid}
     \label{FS_VWEOS_1000x500_RICCA}
\end{figure}

\textcolor{black}{Hence, a supersonic flow of a dense gas with reduced Mach number $M=1.5$ is considered to study the flow over a  forward step} for the test conditions as shown in the table (\ref{dgtable:2Dtransientcases}). Similar to the perfect gas case, a detached bow shock terminates into a Mach reflection compressing the flow into the $\Gamma > 0$ region. The wave centred on the corner of the step takes the form of a physical expansion shock. The flow features obtained are compared to the case as in \cite{argrow}.  Numerical simulations were carried out on two sets of grids $250\times 125$, $500\times 250$ and $1000\times500$ using both MOVERS+ and RICCA algorithms and the fine grid solutions are presented. It can be seen from the figures (\ref{FS_VWEOS_1000x500},\ref{FS_VWEOS_1000x500_RICCA}) that the Mach stem and the reflected shock are well resolved using the numerical schemes.  The change in sign of fundamental derivative is clearly captured by the numerical schemes as shown in figures  (\ref{FS_FD_VWEOS_1000x500},\ref{FS_FD_VWEOS_1000x500_RICCA}).  Further it can be observed from the figures (\ref{FS_DC_VWEOS_1000x500},\ref{FS_DC_VWEOS_1000x500_RICCA}) that the flow structure in the case of dense gas has significant modifications when compared to the perfect gas EOS.


\subsubsection{Supersonic flow over a circular arc}
\begin{figure}[htb!]
\makebox[\textwidth][c]{%
\begin{subfigure}[b]{.4\textwidth}
\centering
\includegraphics[scale=0.1,keepaspectratio]{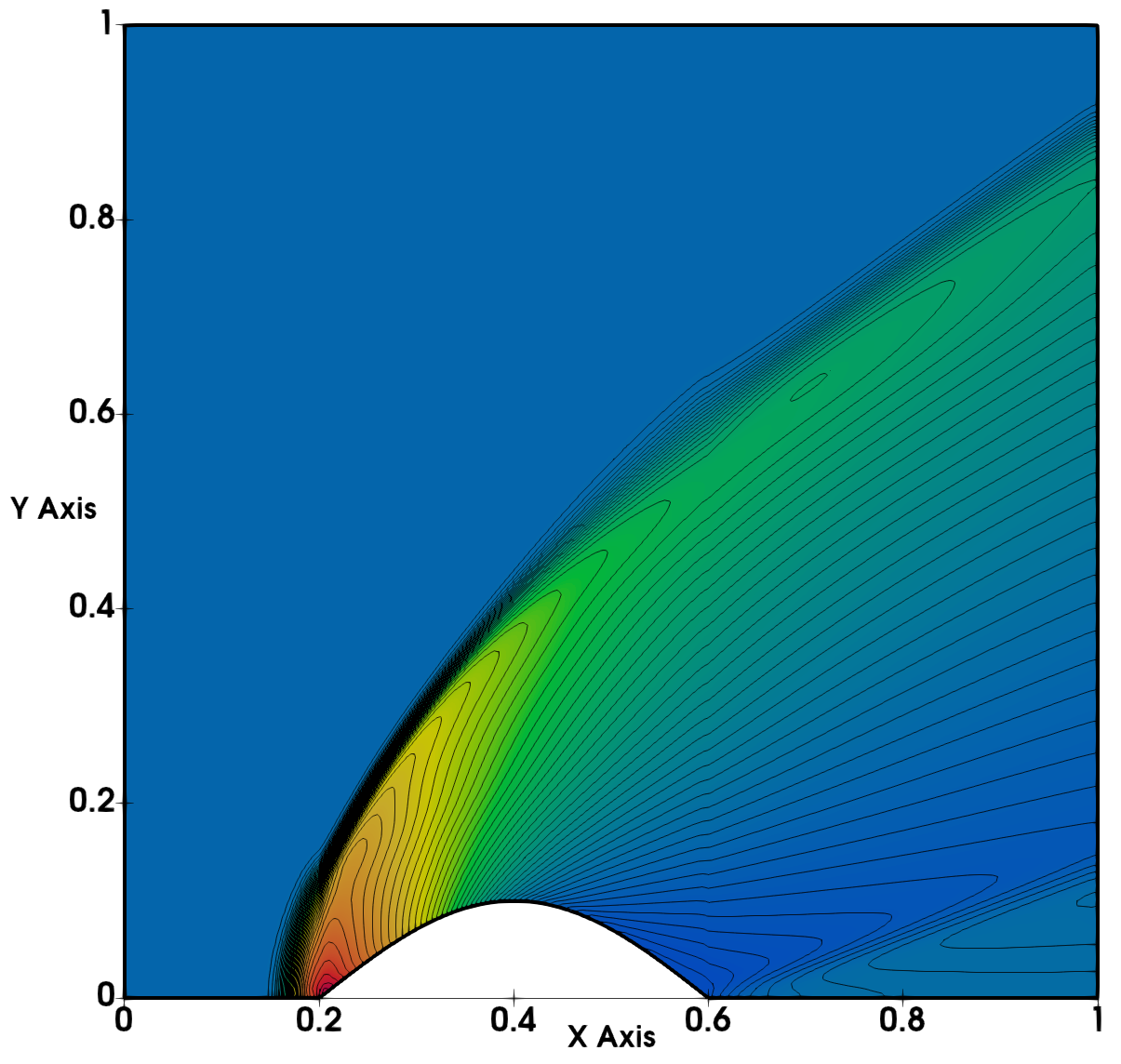}
\caption{Density contours }
\label{FOB_DC_PGEOS}
\end{subfigure}%
\begin{subfigure}[b]{.4\textwidth}
\centering
\includegraphics[scale=0.1,keepaspectratio]{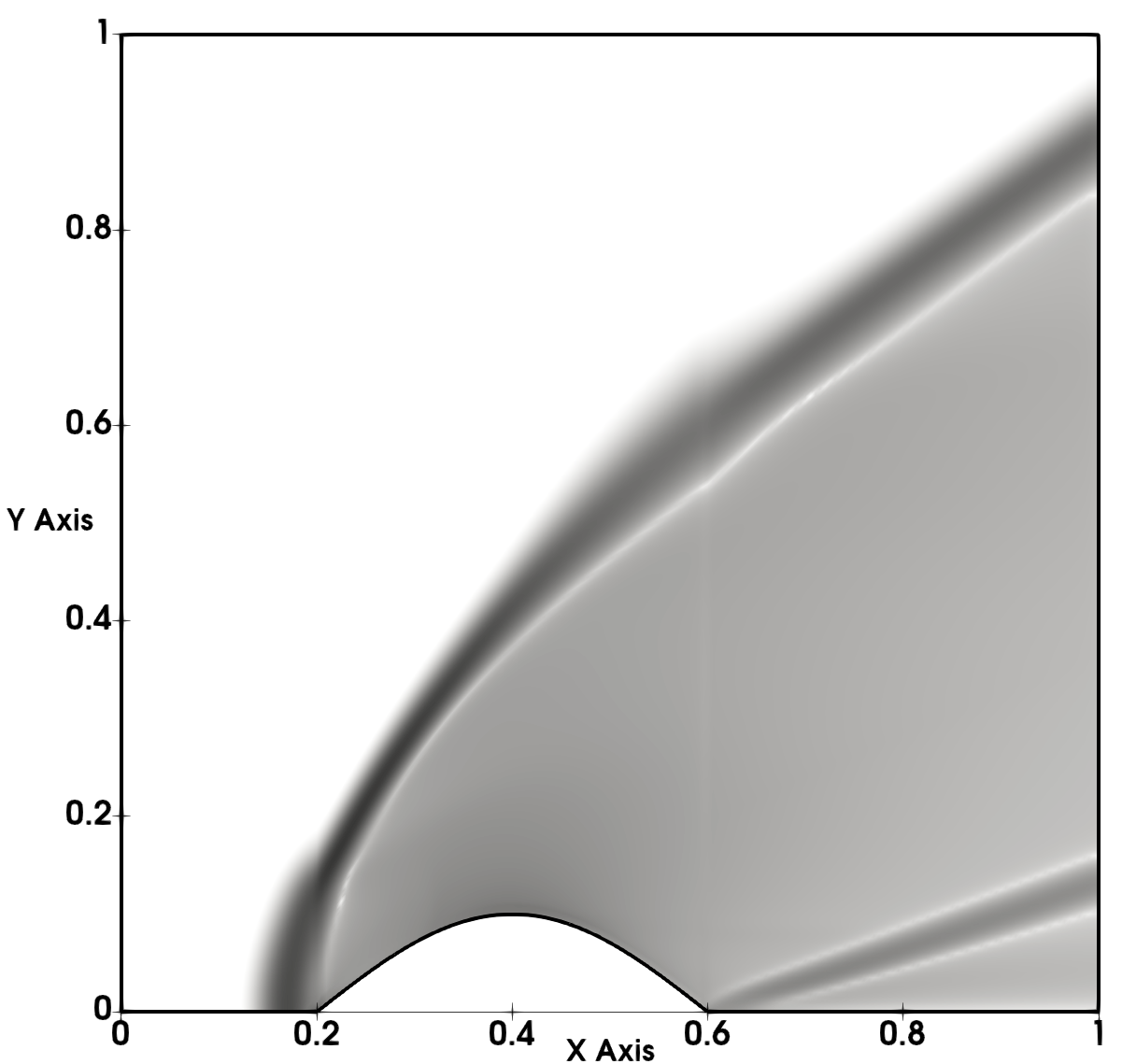}
\caption{Density Gradients}
\label{FOB_DG_PGEOS}
\end{subfigure}
\begin{subfigure}[b]{0.4\textwidth}
\centering
\includegraphics[scale=0.1,keepaspectratio]{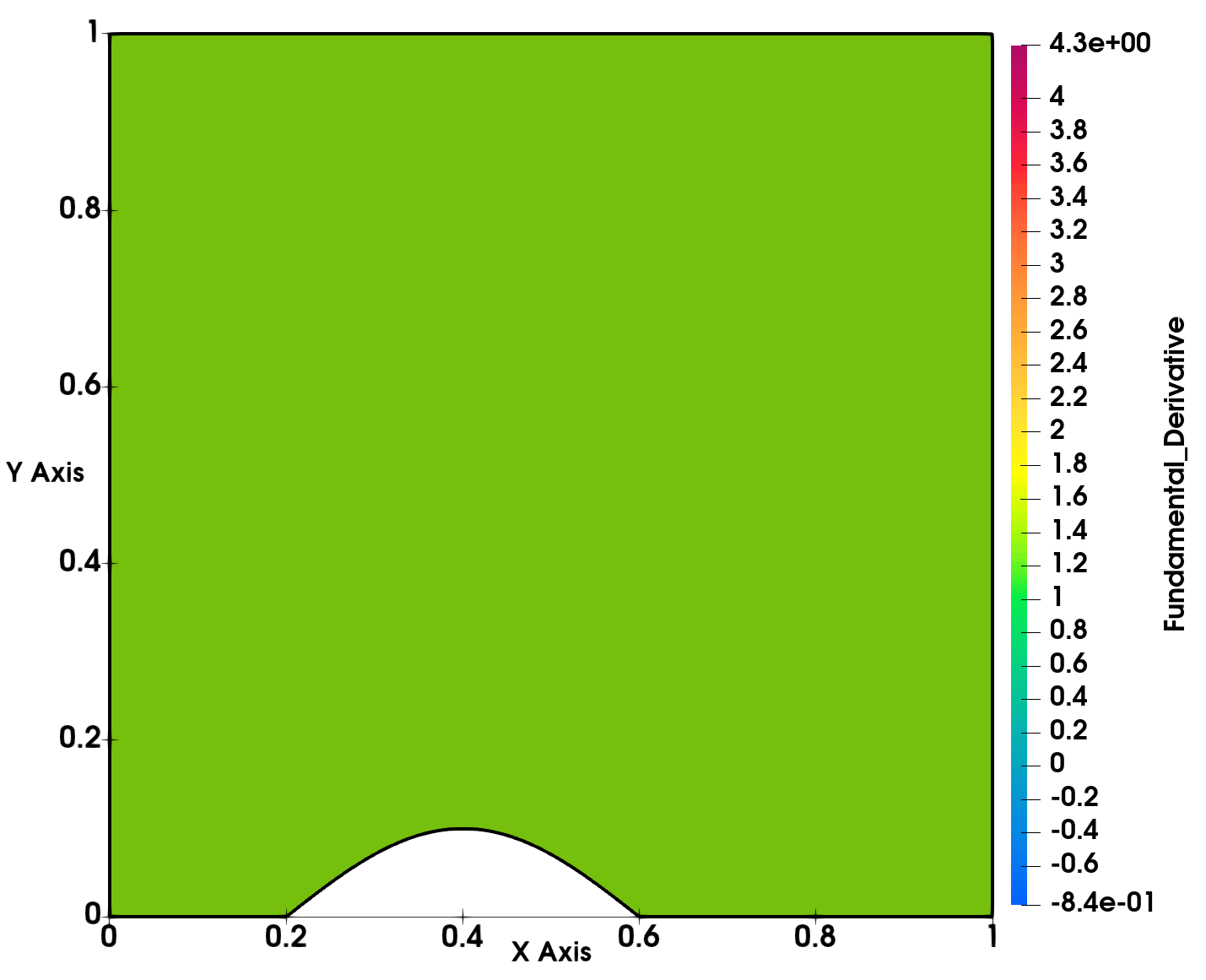}
\caption{Fundamental Derivative}
\label{FOB_FD_PGEOS}
\end{subfigure}

}

\caption{$M=3.0$ flow over circular arc using with Perfect gas EOS}
\label{FOB_PGEOS}
\end{figure}

\begin{figure}[!htb]
\makebox[\textwidth][c]{%
\begin{subfigure}[b]{.4\textwidth}
\centering
 \includegraphics[scale=0.1,keepaspectratio]{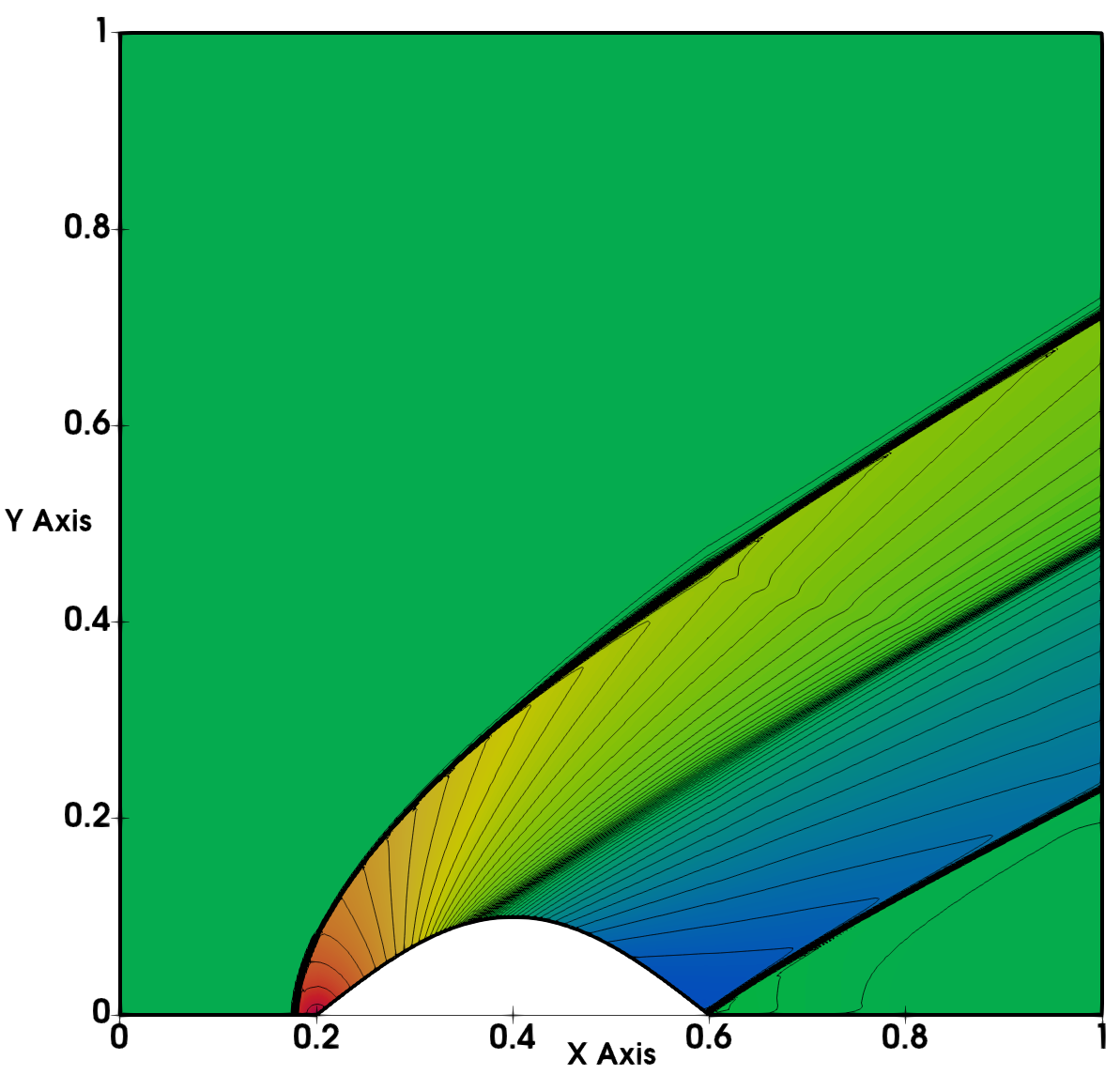}
\caption{Density Contours}
\label{FOB_DC_500x500_MOVERS+}
\end{subfigure}%
\begin{subfigure}[b]{.4\textwidth}
\centering
    \includegraphics[scale=0.1,keepaspectratio]{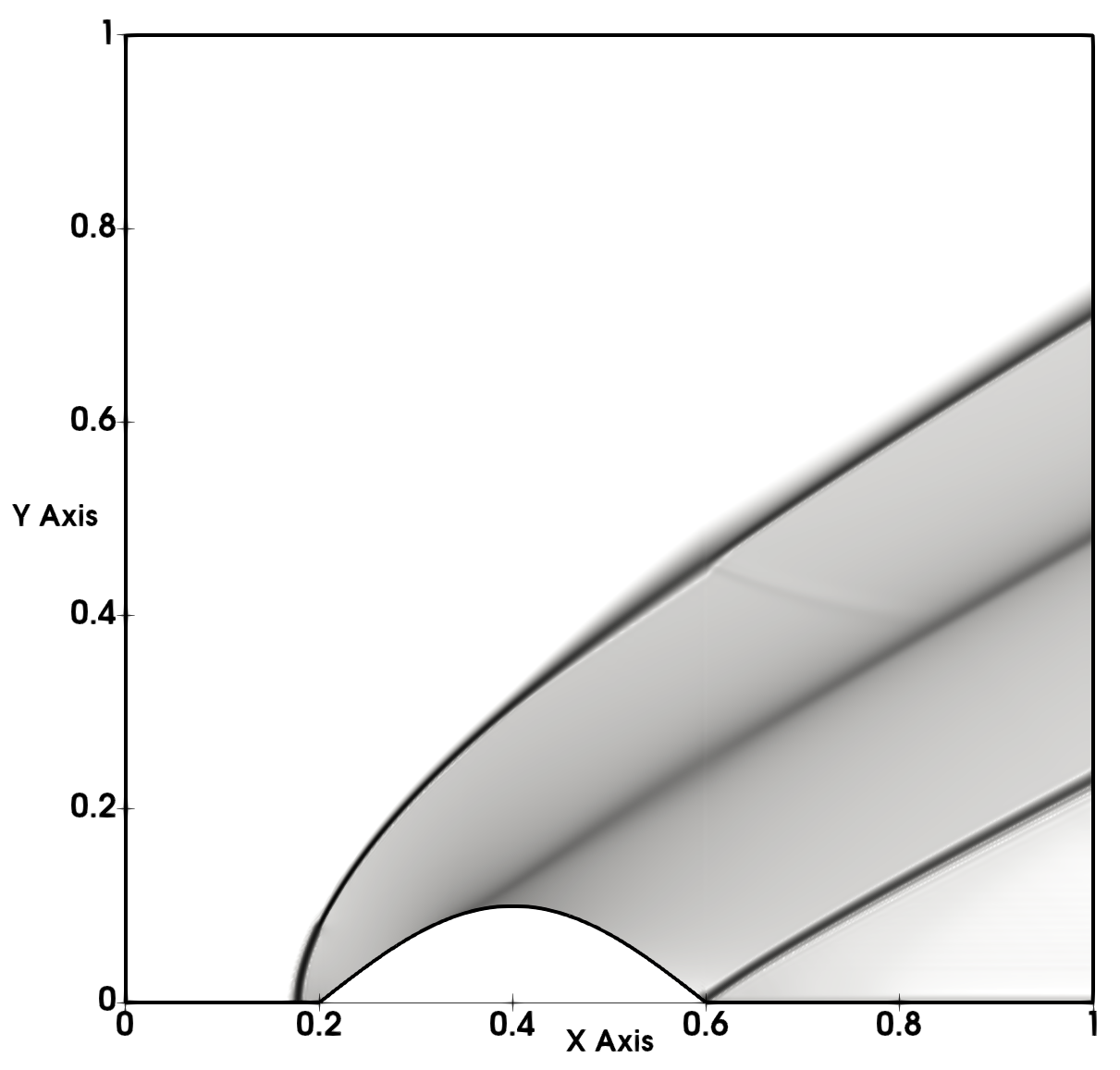}
\caption{Density Gradients}
\label{FOB_DG_500x500_MOVERS+}
\end{subfigure}
\begin{subfigure}[b]{0.4\textwidth}
\centering
    \includegraphics[scale=0.1,keepaspectratio]{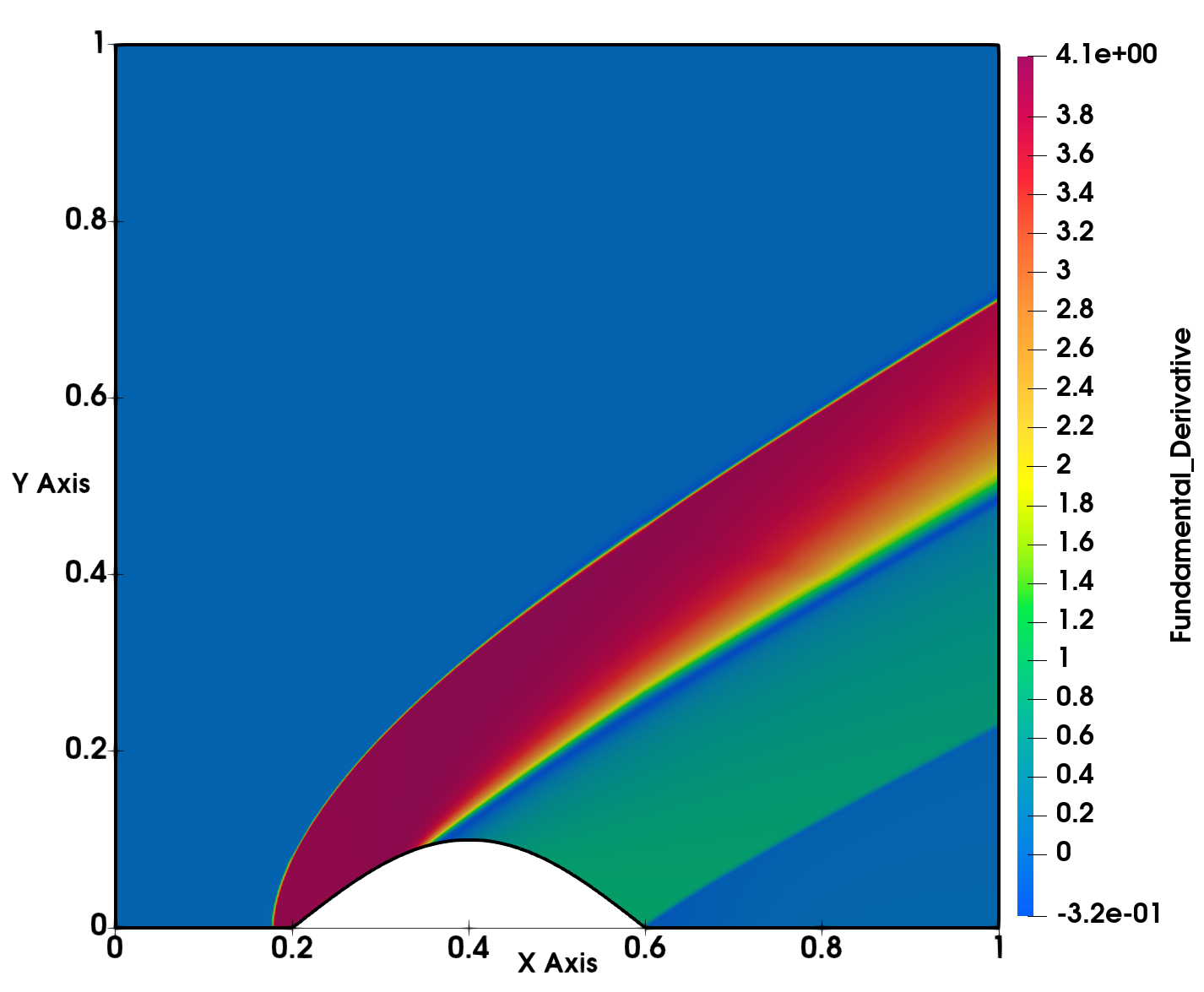}
\caption{Fundamental Derivative}
\label{FOB_FD_500x500_MOVERS+}
\end{subfigure}

}
 \caption{$M=1.5$ Flow over circular arc using Dense Gas with van der Waals EOS using MOVERS+ on $500\times500$ grid} \label{fig:FS_FOB_MOVERS+500x500}
 \end{figure}

\begin{figure}[!htb]
\makebox[\textwidth][c]{%
\begin{subfigure}[b]{.4\textwidth}
\centering
\includegraphics[scale=0.1,keepaspectratio]{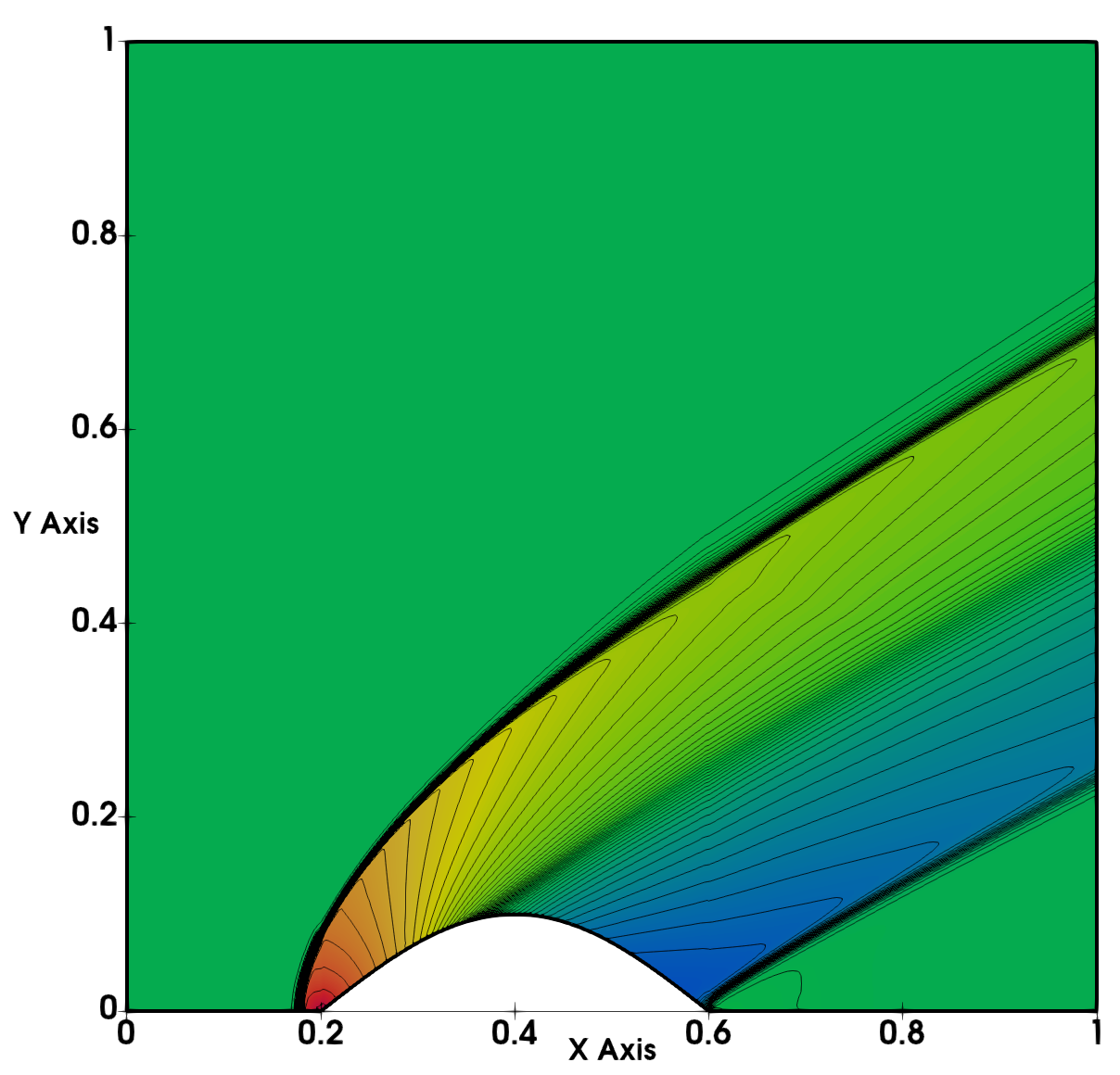}
\caption{Using RICCA}
\label{FOB_DC_500x500_RICCA}
\end{subfigure}%
\begin{subfigure}[b]{.4\textwidth}
\centering
    \includegraphics[scale=0.1,keepaspectratio]{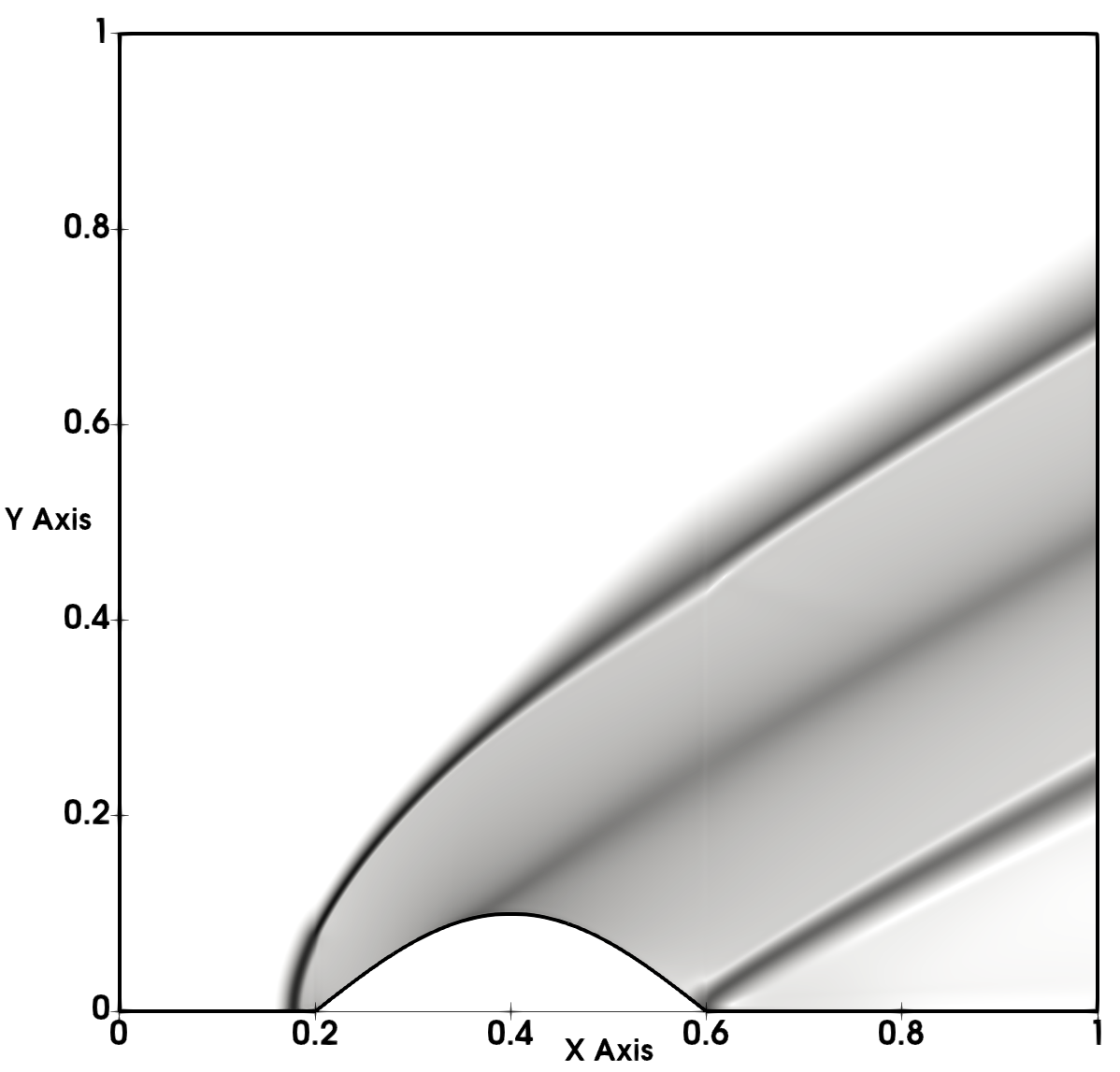}
\caption{Density Gradients}
\label{FOB_DG_500x500_RICCA}
\end{subfigure}
\begin{subfigure}[b]{0.4\textwidth}
\centering
    \includegraphics[scale=0.1,keepaspectratio]{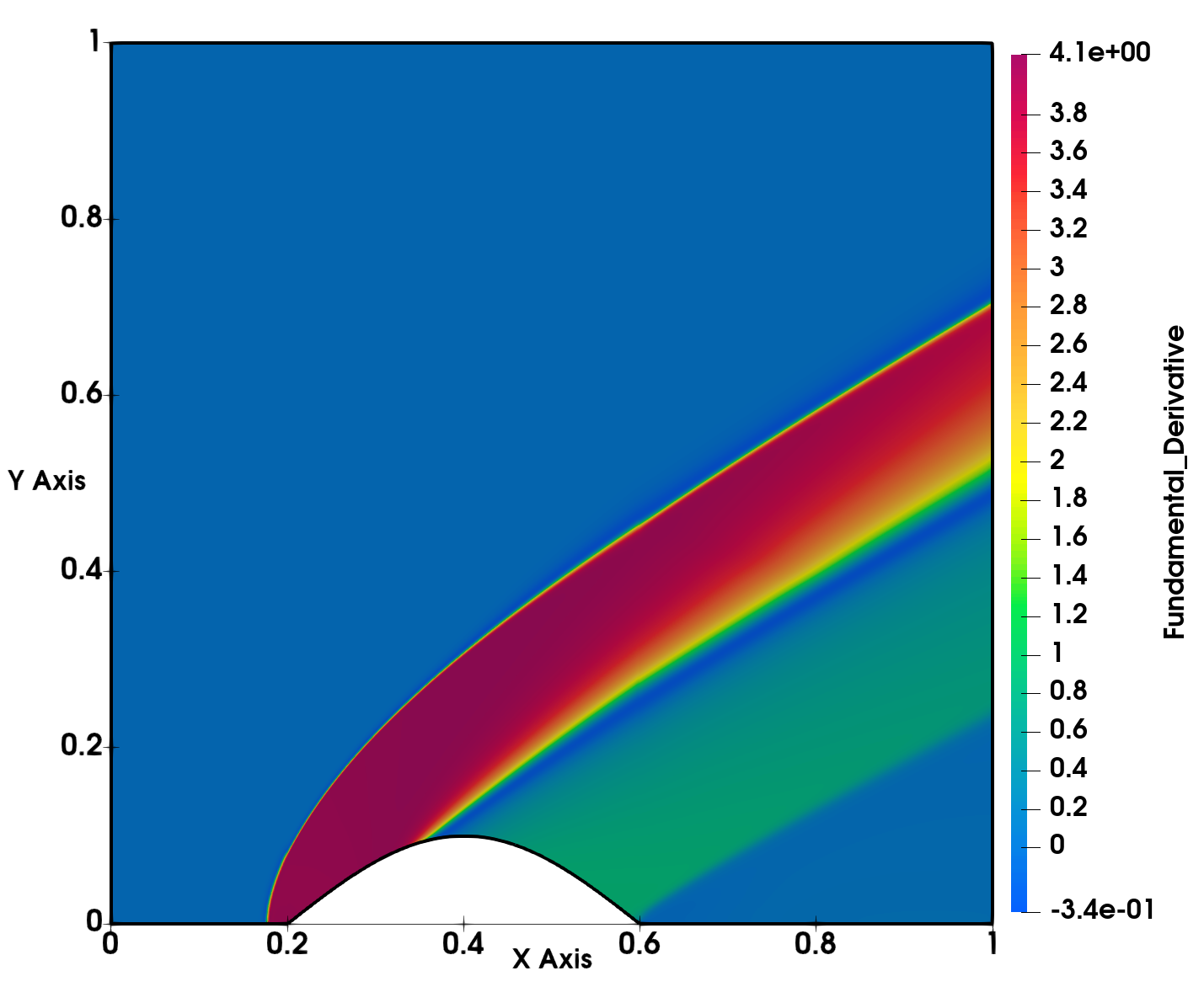}
\caption{Fundamental Derivative}
\label{FOB_FD_500x500_RICCA}
\end{subfigure}

}

\caption{$M=1.5$ Flow over circular arc using Dense Gas with van der Waals EOS using RICCA on $500\times500$ grid} \label{fig:FS_FOB_RICCA500x500}
\end{figure}

 The second test case considered is flow over a circular arc of radius $r = 0.15$, with computational domain $[0,1]\times[0,1]$. A total of $200 \times 200$ control volumes are used for simulation. Consider the flow of a perfect gas case with $M_{\infty} = 3.0$ over this circular bump as shown in figure (\ref{FOB_PGEOS}). The shock is attached to the leading edge of the arc and followed by expansion fan on the curved surface till it encounters a trailing edge shock. It can also be observed from the figure that the fundamental derivative doesn't change its sign across the curved surface.
Next we consider the flow of a dense gas on the circular arc with Mach number $M = 1.5$.  The flow detaches from the leading edge, forming a bow shock. Further the fundamental derivative changes its sign $\Gamma >0$ across the shock and the flow behind the leading-edge forms an expansion shock which expands the flow through $\Gamma <0$ region as explained in \cite{argrow}. Numerical simulations are carried out using MOVERS+ and RICCA using two sets of grids $200\times200$ and $500\times500$. Figures (\ref{FOB_DC_500x500_MOVERS+},\ref{FOB_DC_500x500_RICCA}) represent the density contours on $200\times200$ and $500\times 500$ grids using MOVERS+ and RICCA respectively and fine grid solutions are presented. It can be clearly seen that the flow structure in the expansion region is quite different form the perfect gas solution of figure (\ref{FOB_DC_PGEOS}). It can also be observed that from figures (\ref{FOB_FD_500x500_MOVERS+},\ref{FOB_FD_500x500_RICCA}) the change in the sign of fundamental derivative for dense gas as compared to the fundamental derivative of perfect gas in figure (\ref{FOB_FD_PGEOS}). Both the numerical schemes MOVERS+ and RICCA capture the flow features accurately.

 \subsubsection{Supersonic flow over a smoothly varying expansion ramp}
\begin{figure}[h!]
\makebox[\textwidth][c]{%
\begin{subfigure}[b]{0.5\textwidth}
\centering
 \includegraphics[scale = 0.1,keepaspectratio]{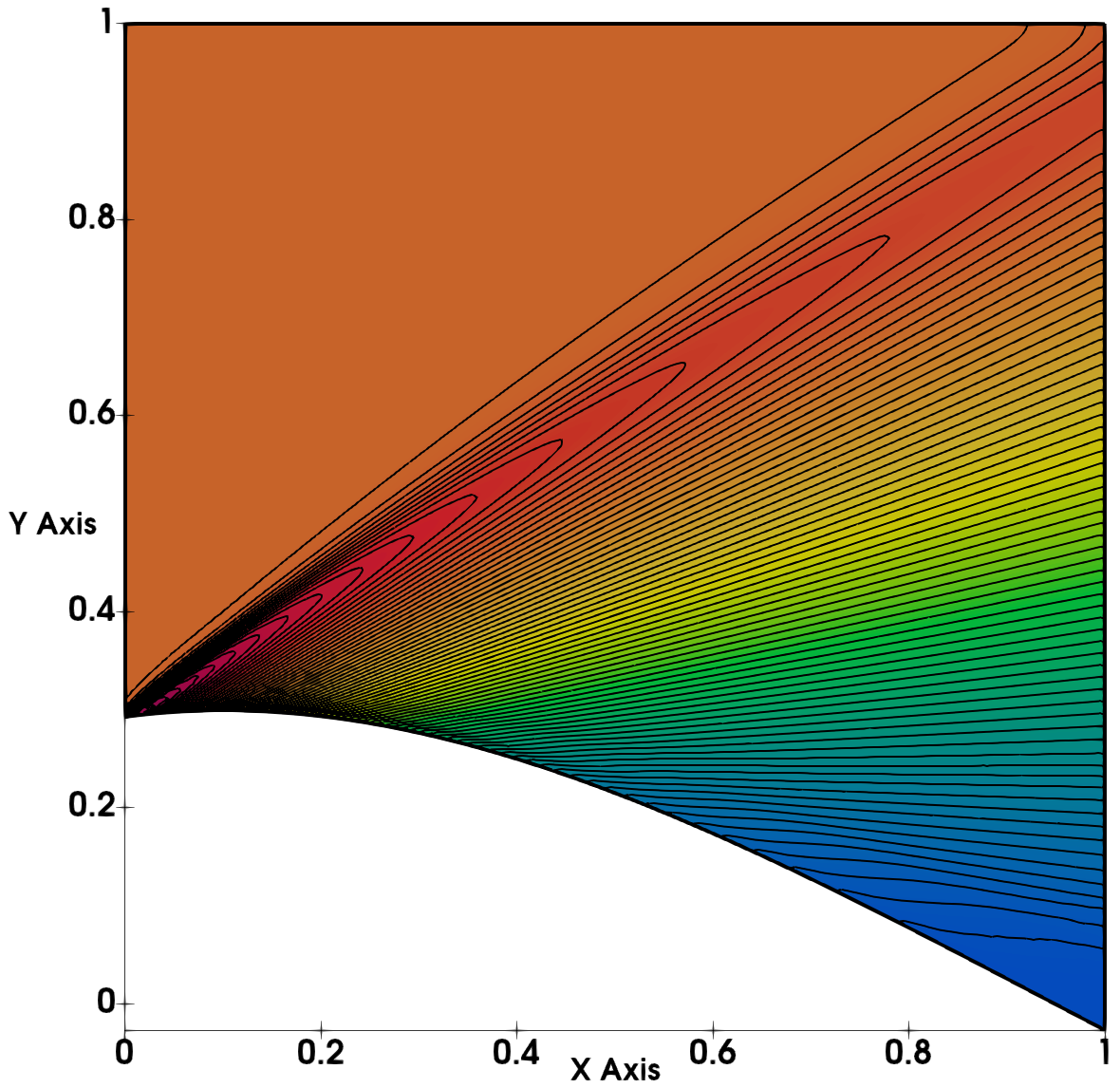}
\caption{Density Contours}
\label{fig:MOVERS+PGEOS_200x200}
\end{subfigure}%
\begin{subfigure}[b]{0.5\textwidth}
\centering
    \includegraphics[scale = 0.1,keepaspectratio]{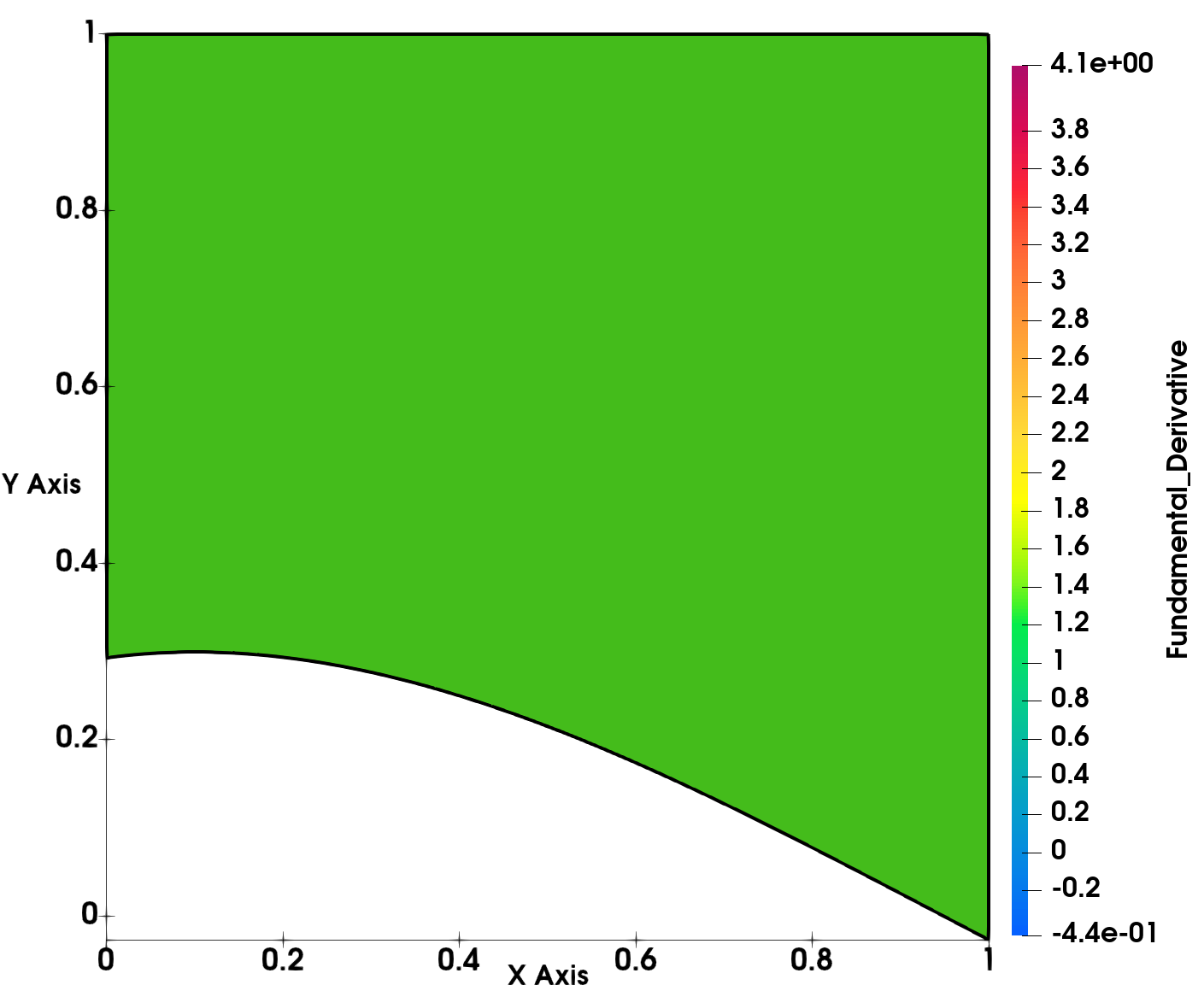}
        \caption{Fundamental Derivative}
\end{subfigure}
}
\end{figure}

In this test case a Mach 2 flow over a smooth expansion ramp with $20^o$ inclination is considered. \textcolor{black}{Three grids are chosen for the study $1000\times1000$,$750\times750$ and $500\times 500 $. Solutions on Fine grid are presented for the purpose. Since the test case is smooth one a grid independent study is also made for this test case.}
 Since the incoming flow is supersonic, it is expected to have a smooth transition and according to the principles of gas dynamics, an expansion fan should evolve from the surface if the flow is for a perfect gas as shown in figure (\ref{fig:MOVERS+PGEOS_200x200}). In the case of dense gas, near the leading edge an expansion shock is observed, as the fundamental derivative transition takes from positive to negative as shown in figures (\ref{movers+500x500}, \ref{RICCA500x500}).
\begin{figure}[htb!]
\makebox[\textwidth][c]{%
\begin{subfigure}[b]{.45\textwidth}
\centering
 \includegraphics[scale = 0.1,keepaspectratio]{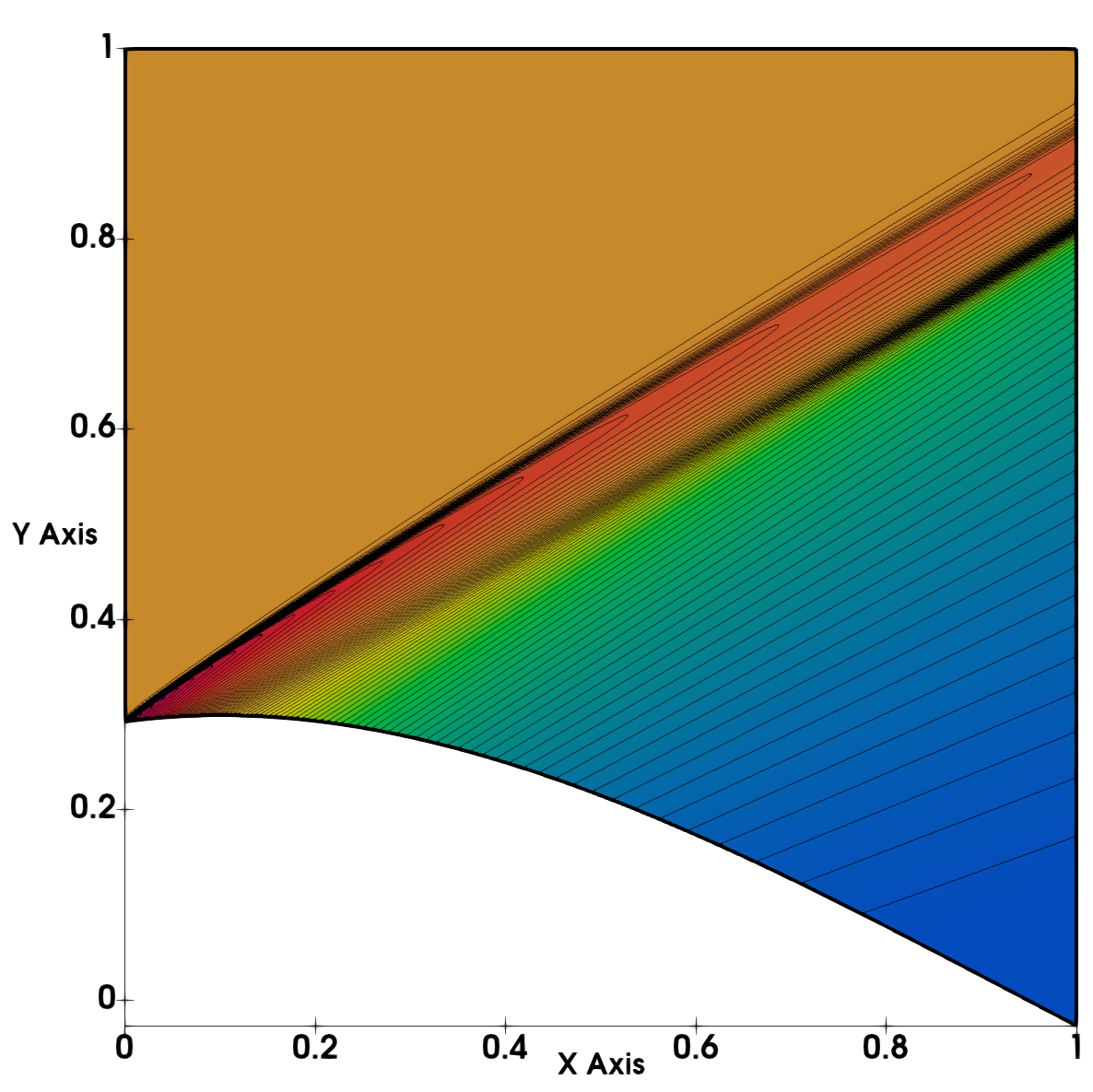}
 \caption{Density Contours}
\end{subfigure}%

\begin{subfigure}[b]{.45\textwidth}
\centering
    \includegraphics[scale = 0.1,keepaspectratio]{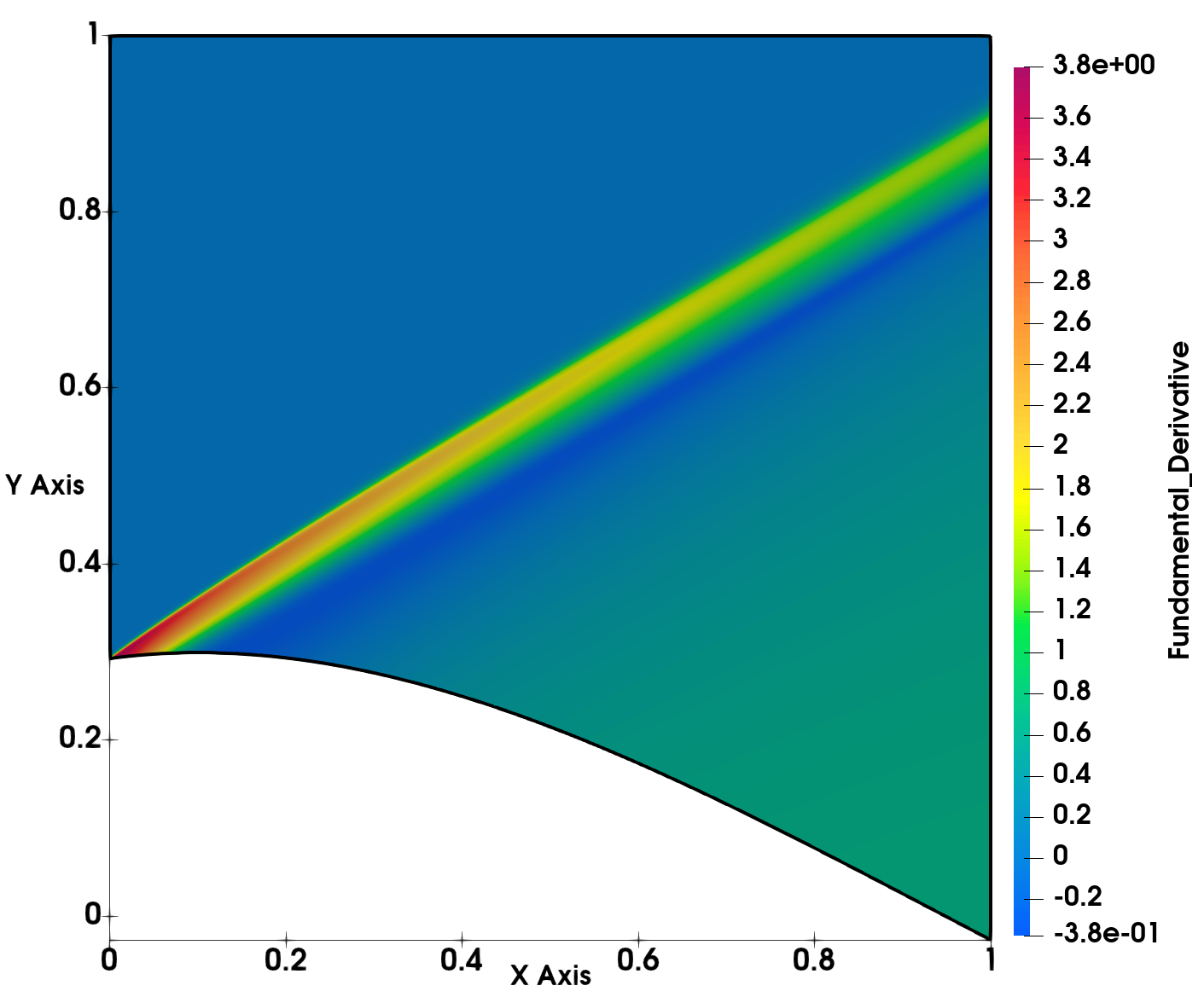}
        \caption{Fundamental Derivative}
\end{subfigure}
}
 \caption{$M=2.0$  over expansion Ramp on $500\times500$ grid using van Der Waals EOS using MOVERS +}
 \label{movers+500x500}
 \end{figure}
 \begin{figure}[htb!]
\makebox[\textwidth][c]{%
\begin{subfigure}[b]{.45\textwidth}
\centering
\includegraphics[scale = 0.1,keepaspectratio]{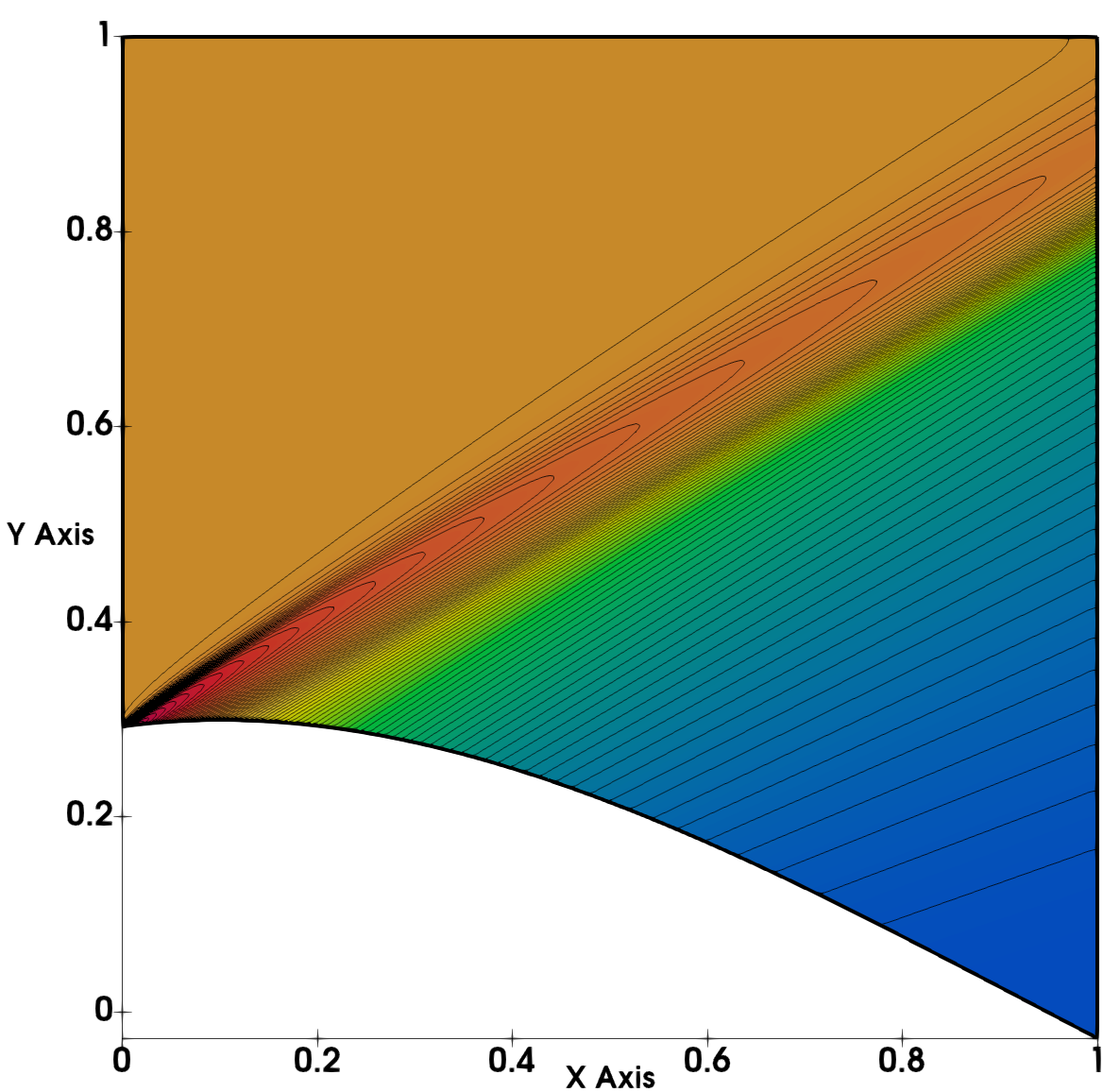}
\caption{Density Contours}
\end{subfigure}%
\begin{subfigure}[b]{.45\textwidth}
\centering
    \includegraphics[scale = 0.1,keepaspectratio]{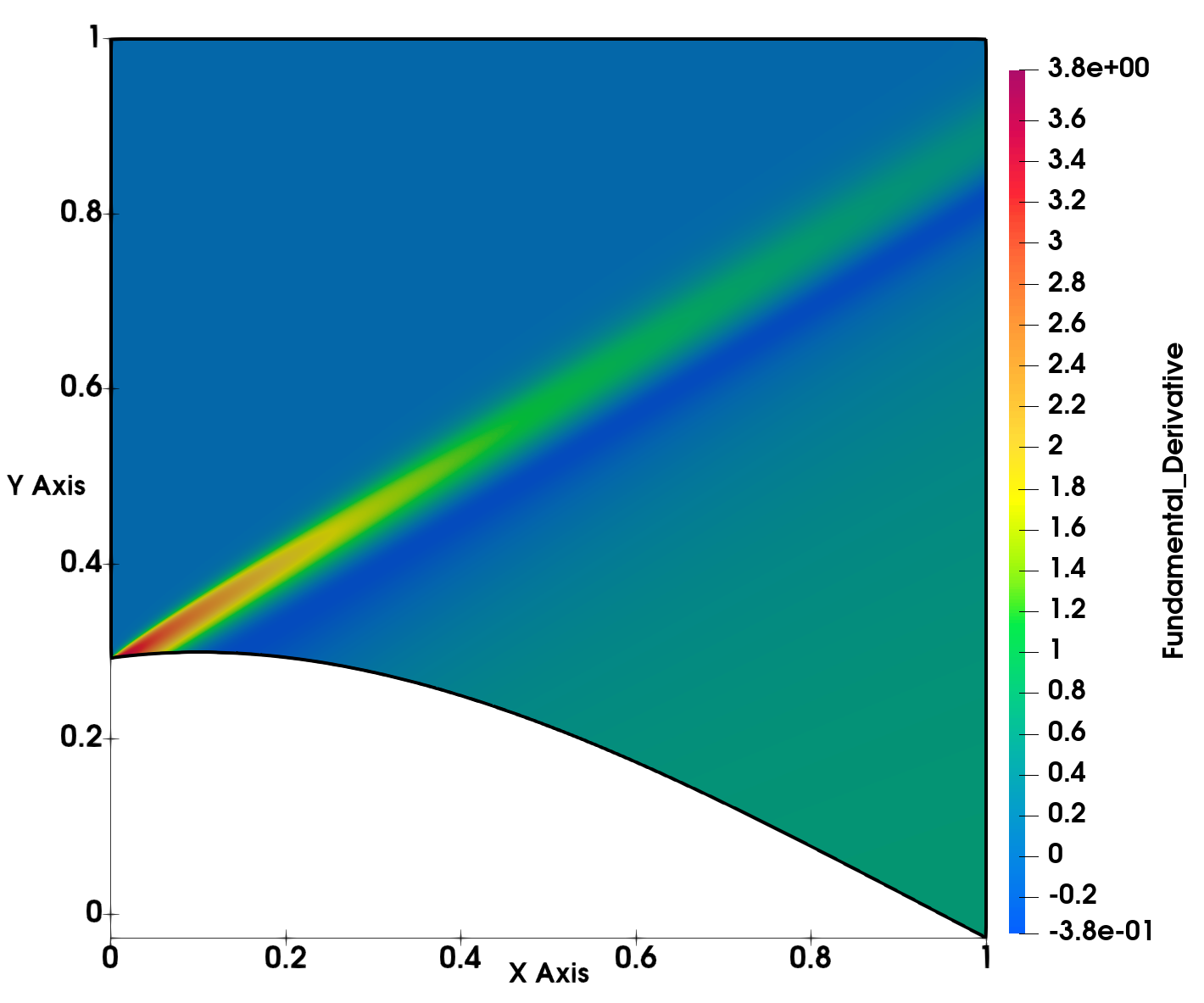}
        \caption{Fundamental Derivative}
\end{subfigure}
}
\caption{$M=2.0$  over expansion Ramp on $500\times500$ grid using van Der Waals EOS using RICCA}
 \label{RICCA500x500}
\end{figure}
\begin{figure}[h!]
\makebox[\textwidth][c]{%
\begin{subfigure}[b]{0.6\textwidth}
\centering
 \includegraphics[scale = 0.15,keepaspectratio]{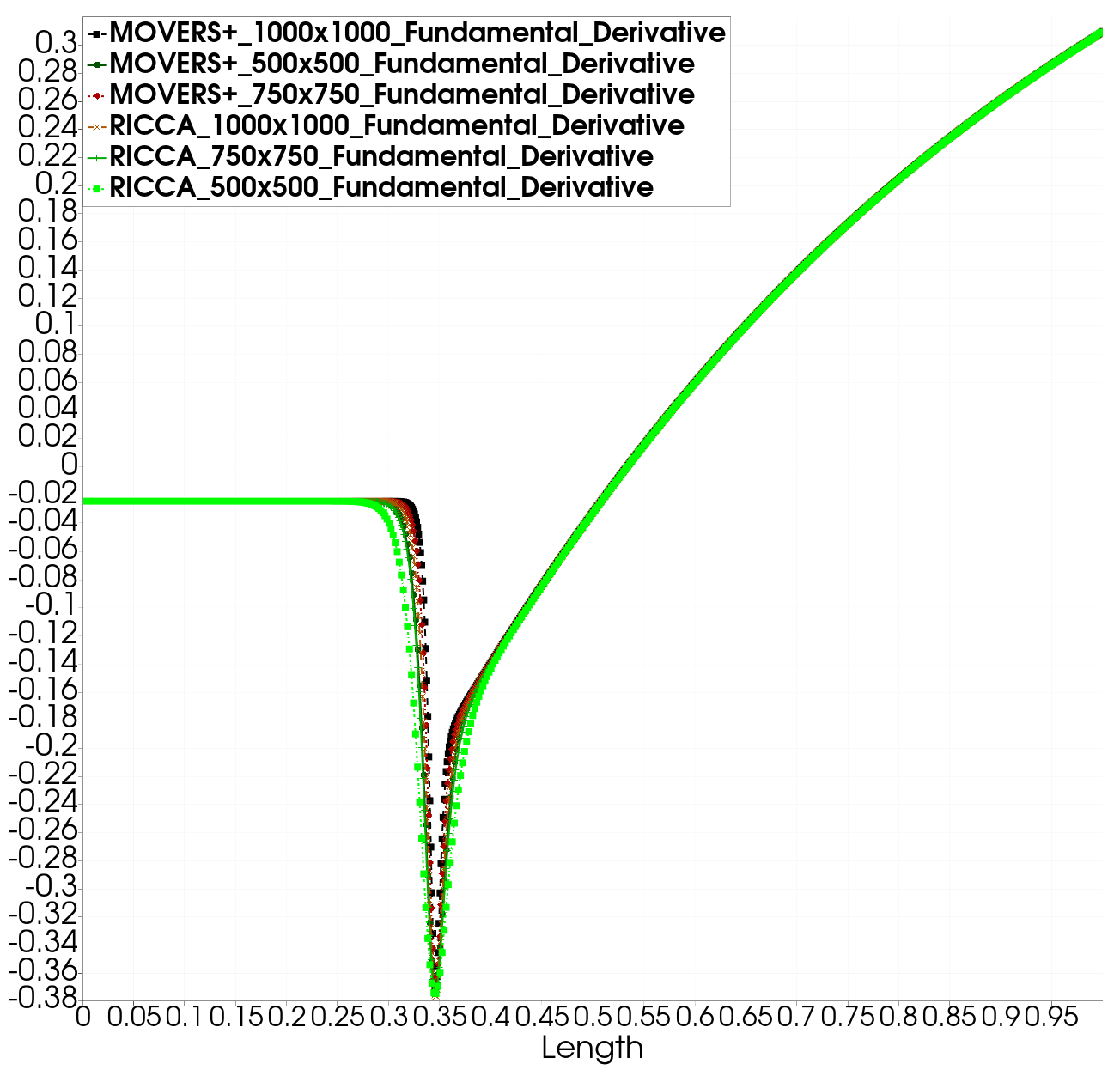}
 \caption{Fundamental derivative}
\end{subfigure}%
\begin{subfigure}[b]{0.6\textwidth}
\centering
    \includegraphics[scale = 0.15,keepaspectratio]{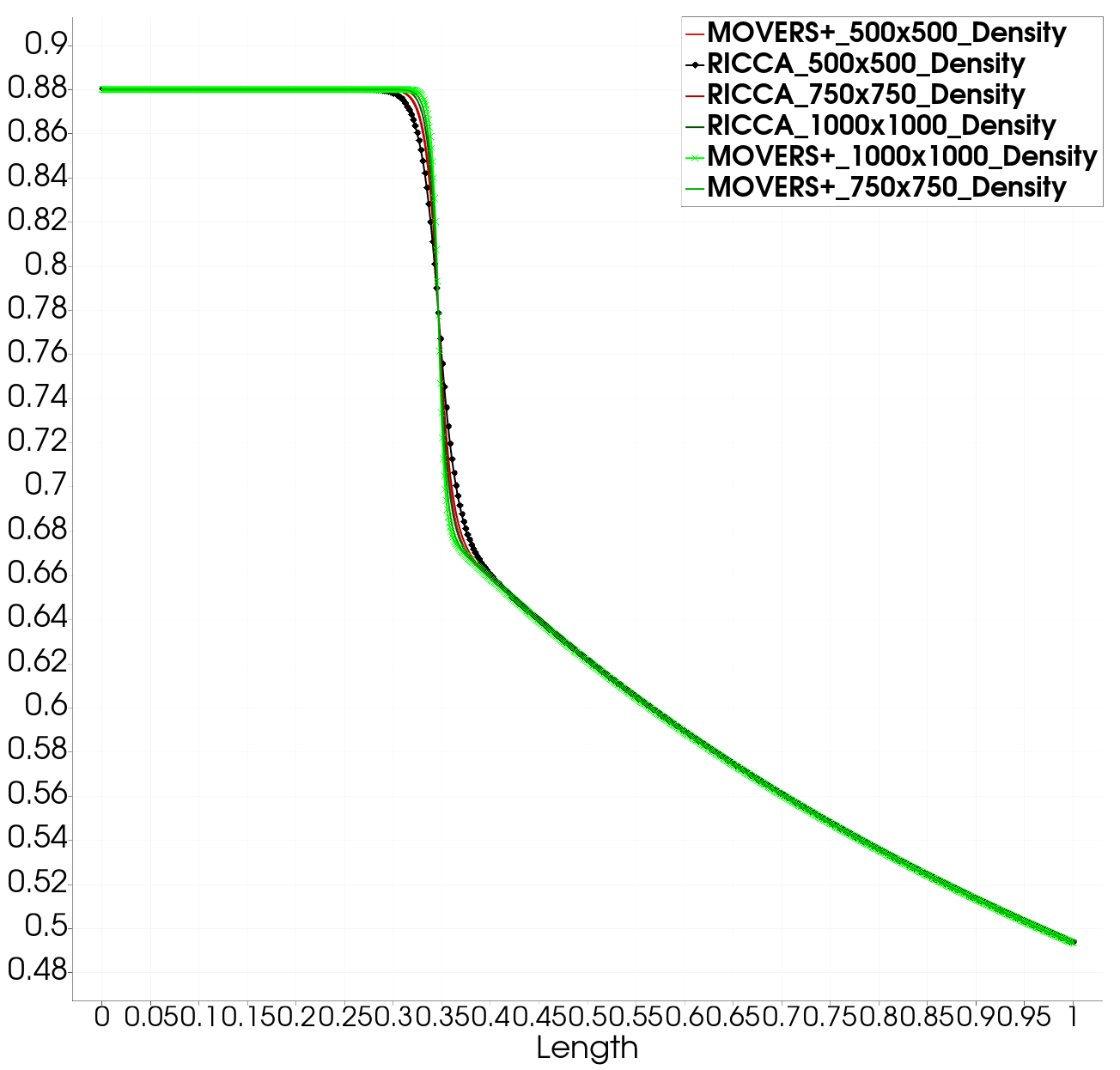}
        \caption{Density}
\end{subfigure}
}
 \caption{Grid Independence study over expansion Ramp using MOVERS+ and RICCA}
 \label{GridIndep}
 
\end{figure}
\subsubsection{Transient test cases - Supersonic flow over backward facing step}
\textcolor{black}{The transient test case considered here is a moving shock wave refracting over a backward facing step, also known as shock diffraction test case. Initial conditions for this test case are given in table (\ref{dgtable:2Dtransientcases}). Transient conditions for test cases using perfect gas EOS are indicated as TPG similarly transient cases for dense gases are indicated as TDG, where 1 and 2  on variables refer to pre-shock and post-shock conditions. Since this is a transient case, for resolving time in the numerical simulation a third order Range Kutta method is used for discretisation of time derivative.} 
 \begin{table}[htb!]
\centering
\begin{tabular}{ |c|c|c|c|c|c|c|}
\hline
Case & $p_2$ & $\rho_2$ & $u_2$ & $p_1$& $\rho_2$ & $u_1$\\
\hline
TPG&12.164&3.06&2.74&1.00&1.00&0.0\\
\hline
TDG&0.98&0.62&-0.14&1.09&0.88&0.0\\
\hline
\end{tabular}
\caption{Initial conditions for transient test cases for simulation of perfect gas and dense gases}
\label{dgtable:2Dtransientcases}
\end{table}

\begin{figure}[htb!]
\makebox[\textwidth][c]{%
\begin{subfigure}[b]{.6\textwidth}
\centering
 \includegraphics[scale = 0.15,keepaspectratio]{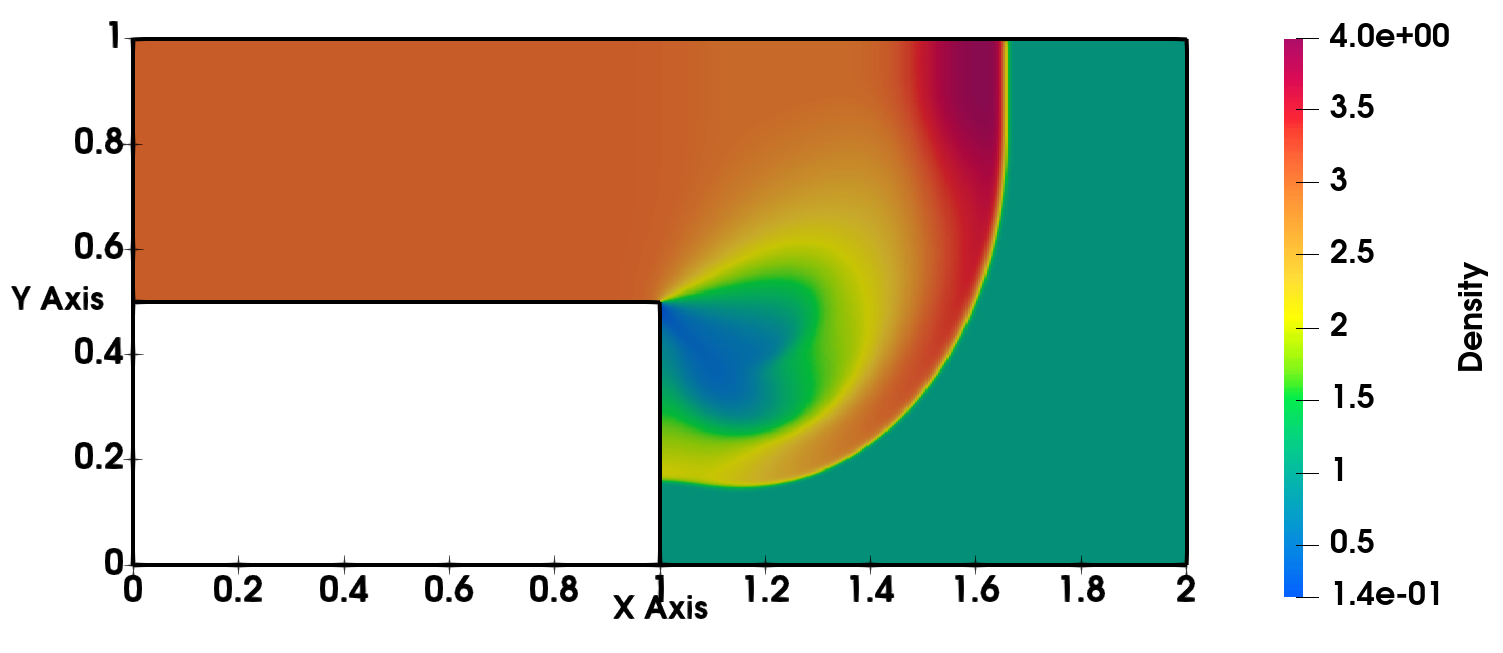}
 \caption{Density Contours}
 \label{DC_500x500}
\end{subfigure}%
\begin{subfigure}[b]{.6\textwidth}
\centering
    \includegraphics[scale = 0.15,keepaspectratio]{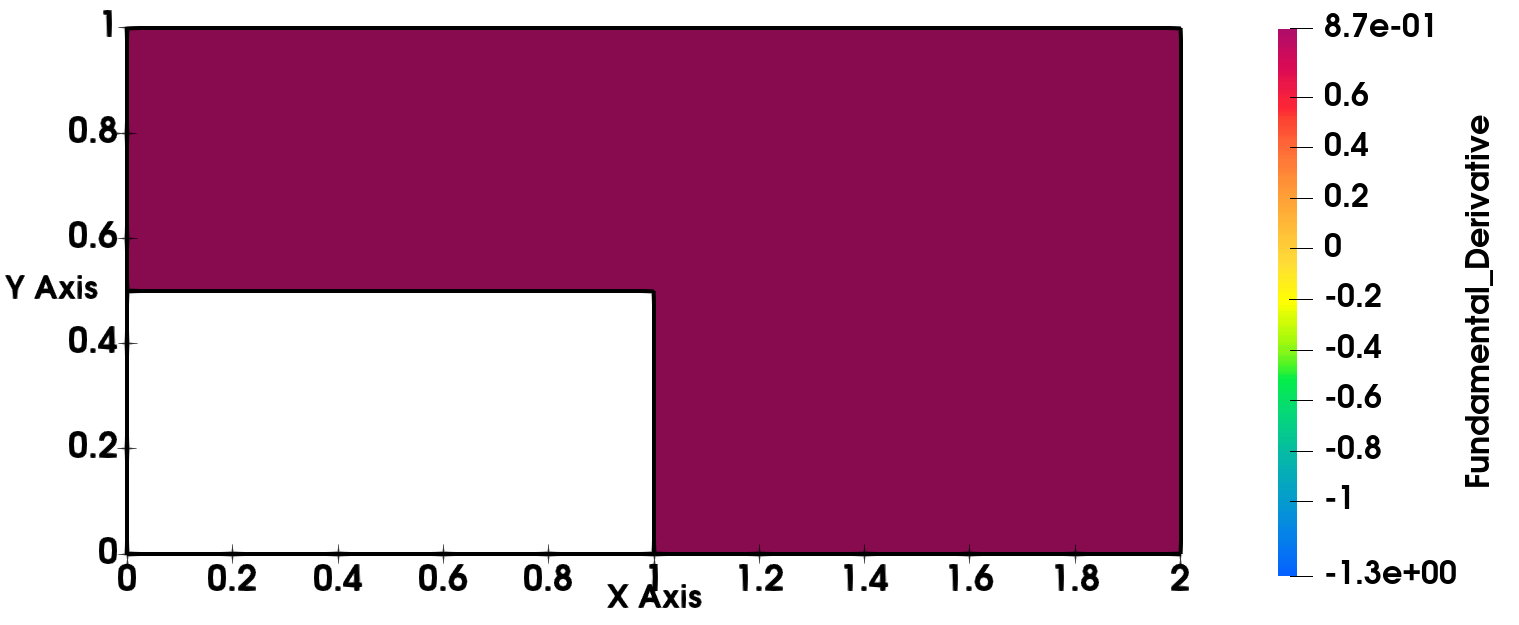}
        \caption{Fundamental Derivative }
 \label{FD_500x500}  
 \end{subfigure}%
}
       \caption{Shock Diffraction with Perfect gas EOS}       
\end{figure}

\textcolor{black}{The solution of perfect gas consists of an expansion fan emanating from the corner, a contact discontinuity and a propagating shock wave, further the fundamental derivative doesn't change in the domain. Where as for a dense gas, instead of an expansion fan at the corner of the backward facing step an expansion shock evolves because the fundamental derivative changes it sign from negative to positive.} 
\textcolor{black}{The solution with perfect gas EOS are presented in figure(\ref{DC_500x500}). It can also be seen that the fundamental derivative doesn't change its sign as seen from the  figure(\ref{FD_500x500}).}

\begin{figure}[htb!]
\makebox[\textwidth][c]{%

\begin{subfigure}[b]{.6\textwidth}
\centering
 \includegraphics[scale = 0.15,keepaspectratio]{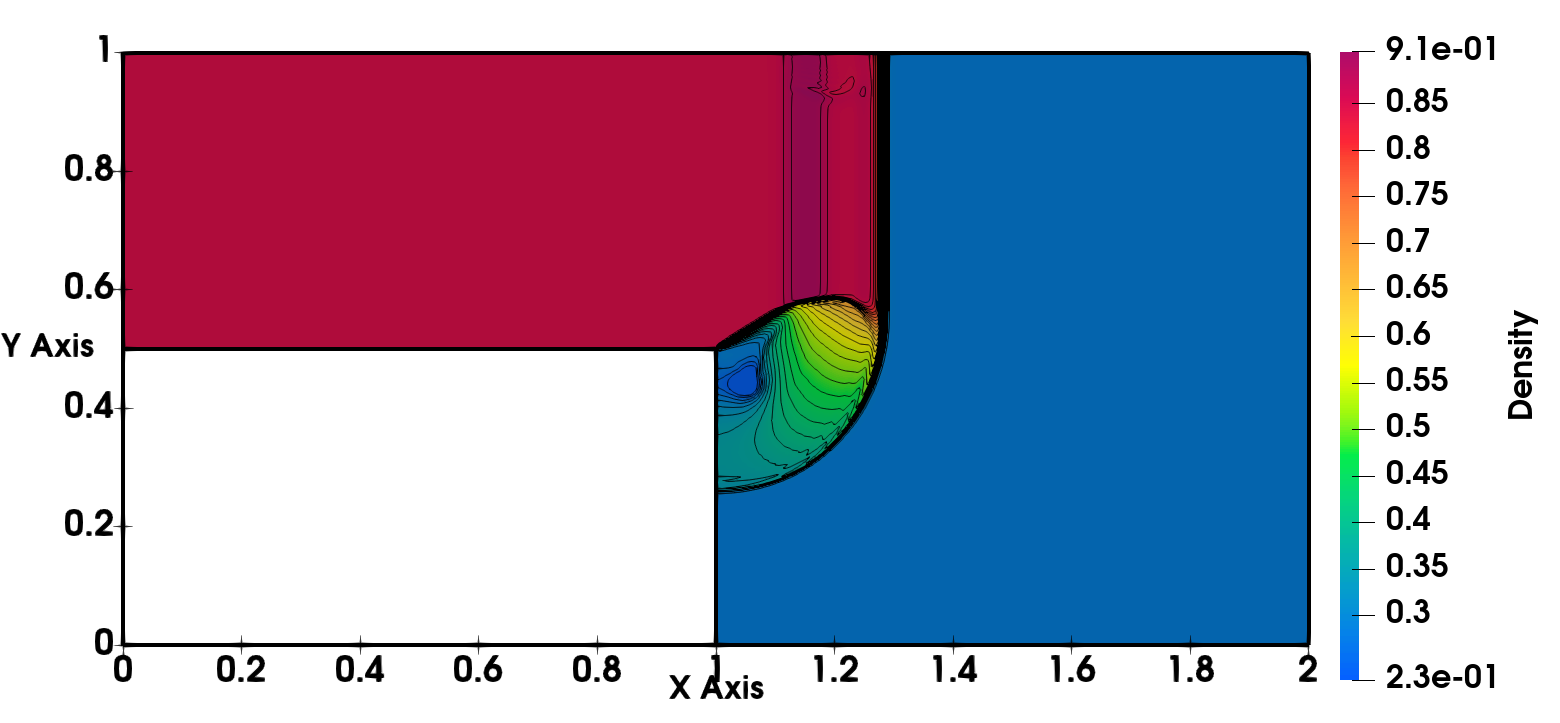}
 \caption{at t = 0.3}
 \label{DC_Movers+_751x751_t03}
\end{subfigure}%

\begin{subfigure}[b]{.6\textwidth}
\centering
    \includegraphics[scale = 0.15,keepaspectratio]{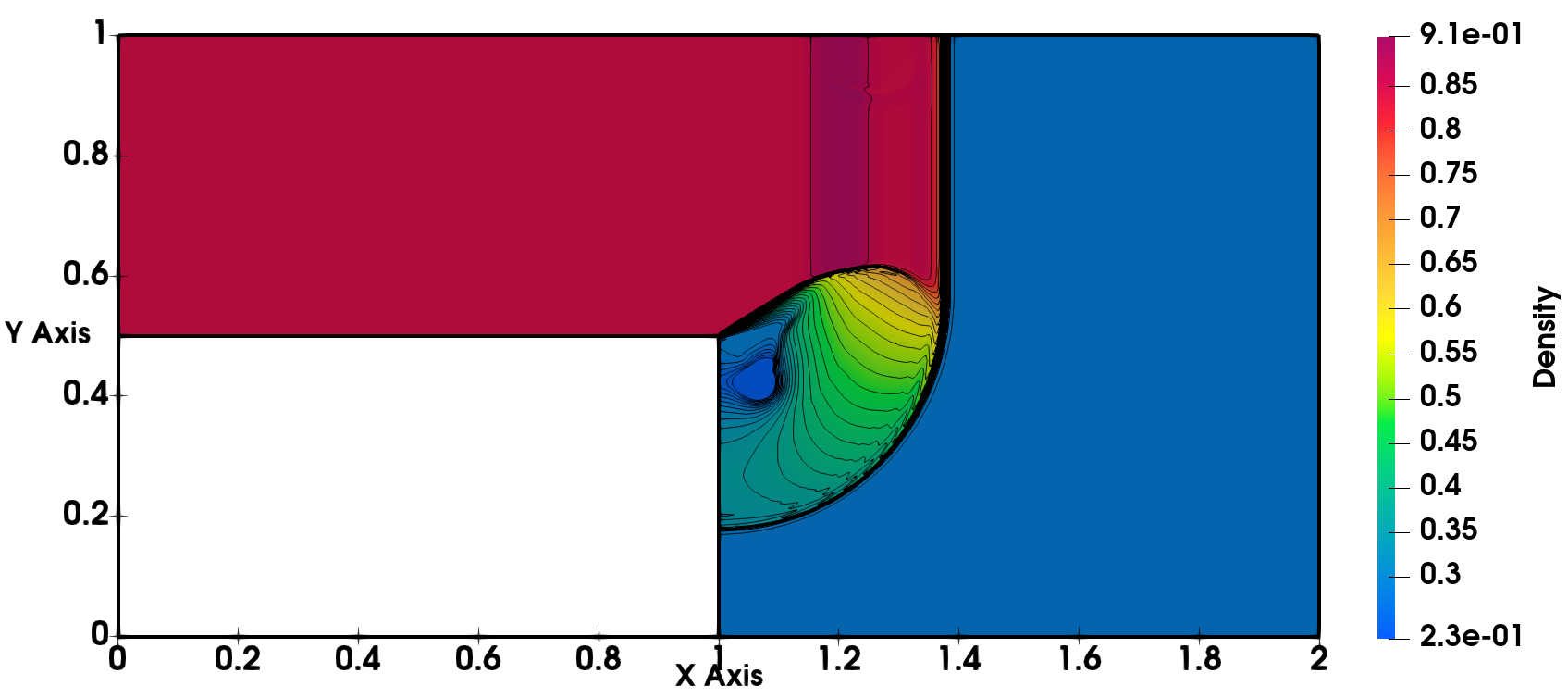}
        \caption{at t = 0.4}
         \label{DC_Movers+_751x751_t04}
\end{subfigure}
}
%
\makebox[\textwidth][c]{%
\begin{subfigure}[b]{0.6\textwidth}
\centering
    \includegraphics[scale = 0.15,keepaspectratio]{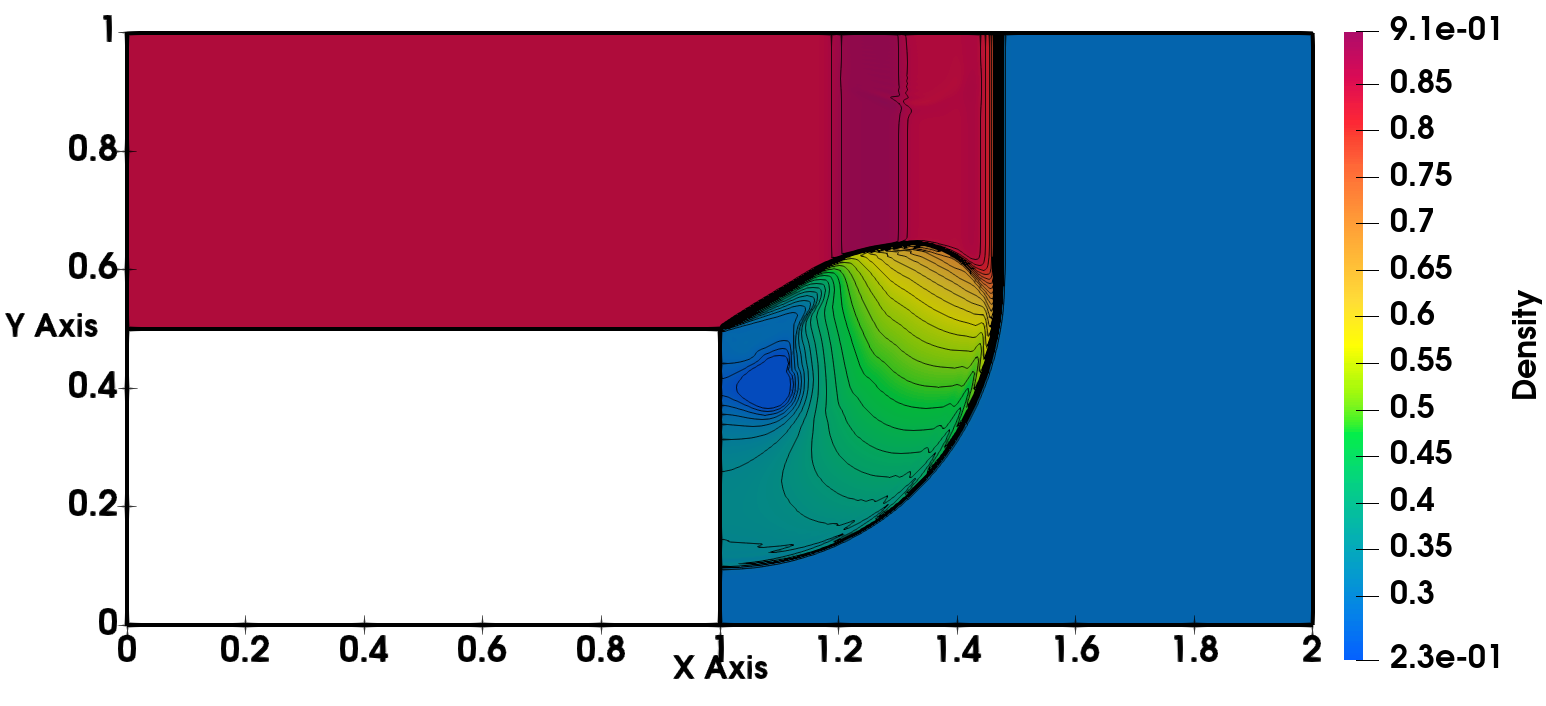}
        \caption{at t = 0.5}
         \label{DC_Movers+_751x751_t05}
\end{subfigure}
\begin{subfigure}[b]{0.6\textwidth}
\centering
\includegraphics[scale = 0.15,keepaspectratio]{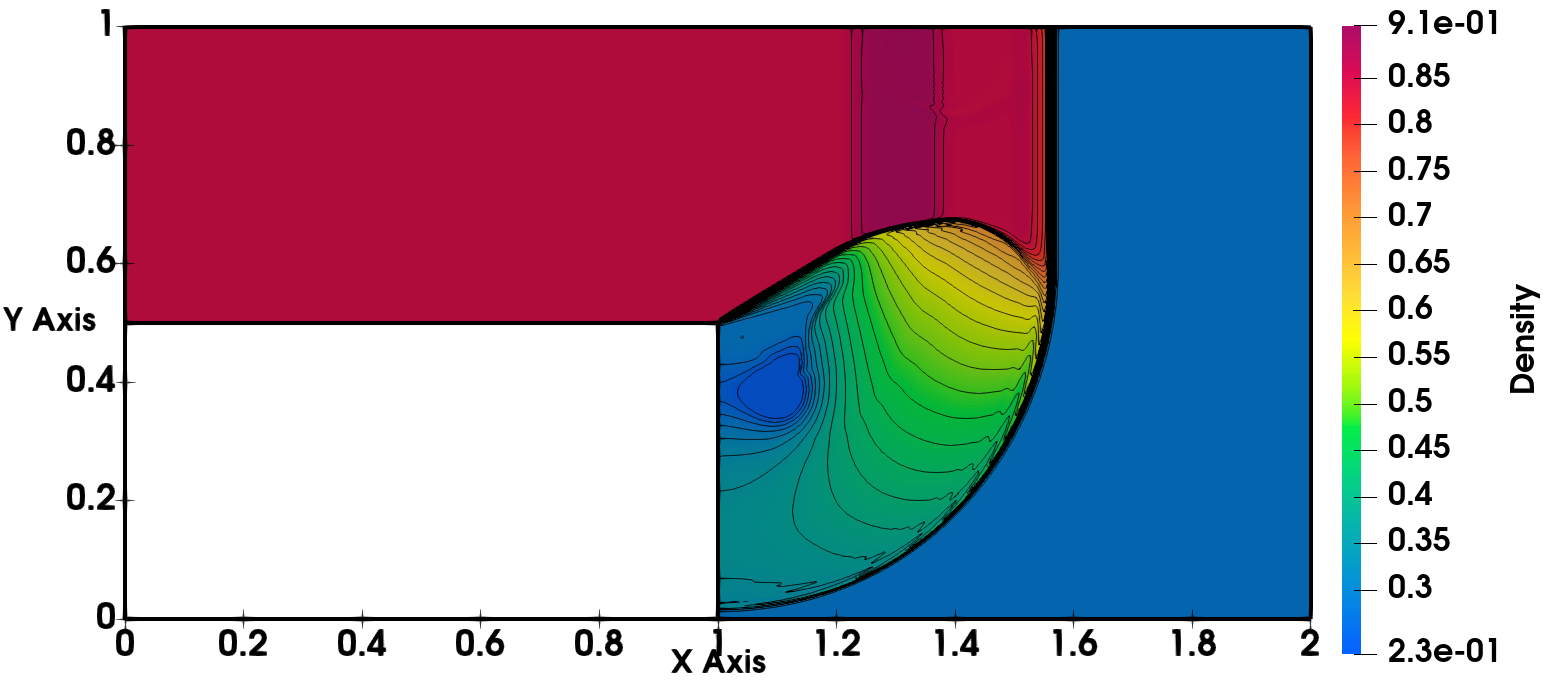}
\caption{t=0.6}
 \label{DC_Movers+_751x751_t06}
\end{subfigure}%
}
 \caption{Shock Diffraction with $M=1.23$  on $751\times751$ grid using van Der Waals EOS and MOVERS + at various times}
 \label{movers+751x751_varioust}
 \end{figure}

\textcolor{black}{Numerical simulations are carried out using MOVERS+ and RICCA on $500\times500$, $750\times750$ and $1000\times1000$ grid sizes to observe the flow features and also to study grid independence. Results of simulation are presented at $t = 0.4$ on $500\times500$ and $1000\times1000$ grid and to check if there exists any instability in the numerical scheme, simulations are extended for further time on all the grids and the solution is presented on $750\times750$ grid as shown in figures (\ref{DC_Movers+_751x751_t03},\ref{DC_Movers+_751x751_t04},\ref{DC_Movers+_751x751_t05},\ref{DC_Movers+_751x751_t06}).}

\begin{figure}[htb!]
\makebox[\textwidth][c]{%
\begin{subfigure}[b]{0.6\textwidth}
\centering
 \includegraphics[scale = 0.15,keepaspectratio]{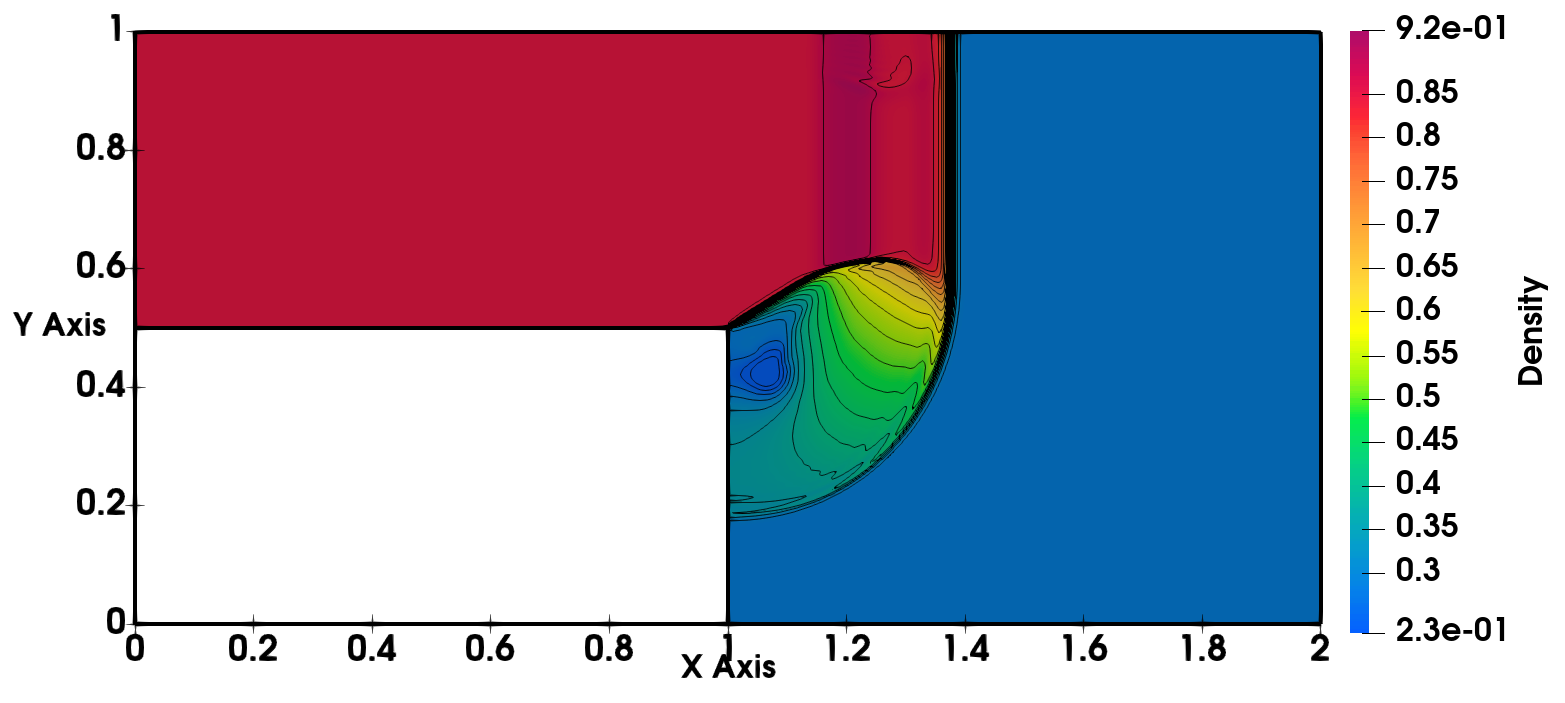}
 \caption{Density Contours}
 \label{DC_Movers+_500x500}
\end{subfigure}%
\begin{subfigure}[b]{0.6\textwidth}
\centering
    \includegraphics[scale = 0.15,keepaspectratio]{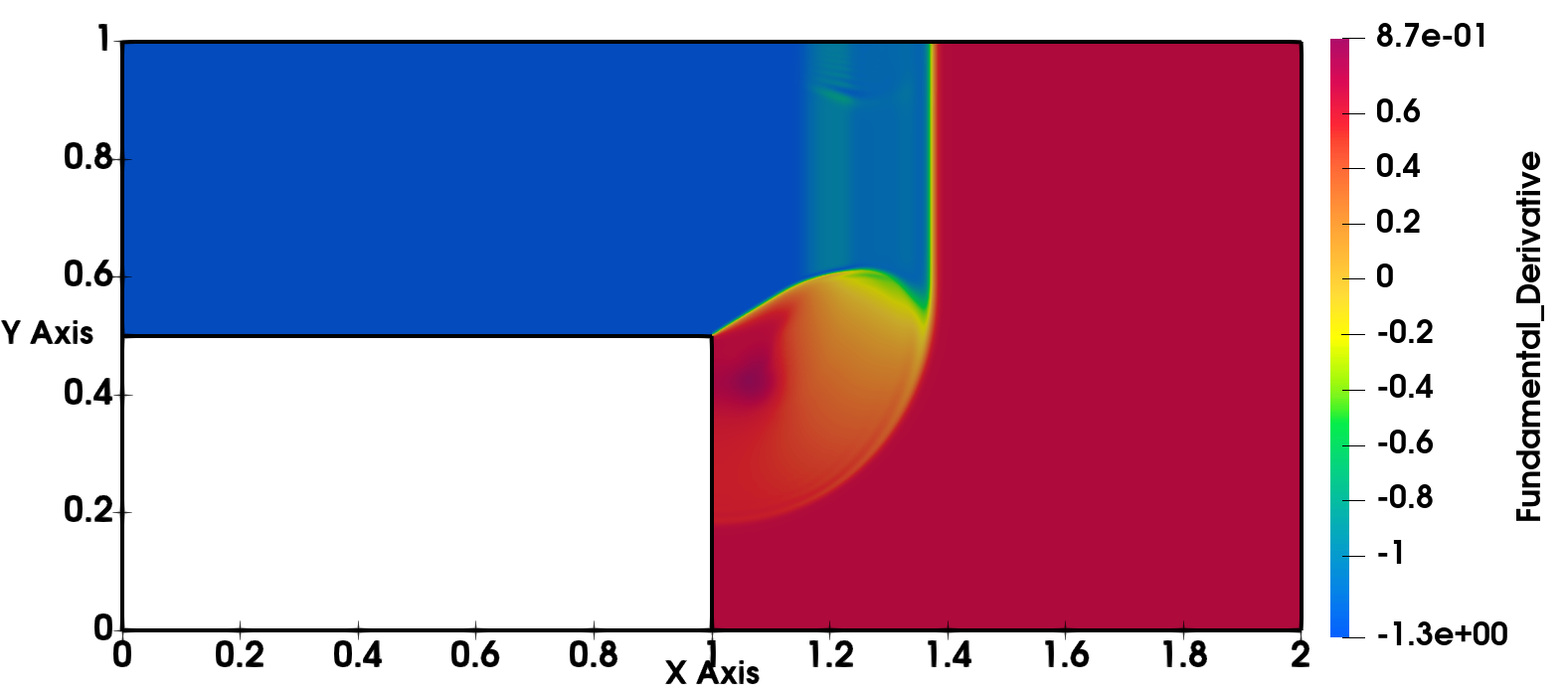}
        \caption{Fundamental Derivative }
 \label{FD_Movers+_500x500}        
\end{subfigure}
}
 \caption{Shock Diffraction with $M=1.23$  on $500\times500$ grid using van Der Waals EOS and MOVERS + at t = 0.4}
 \label{movers+500x500}
  \end{figure}

\begin{figure}[htb!]
\makebox[\textwidth][c]{%
\begin{subfigure}[b]{0.6\textwidth}
\centering
\includegraphics[scale = 0.15,keepaspectratio]{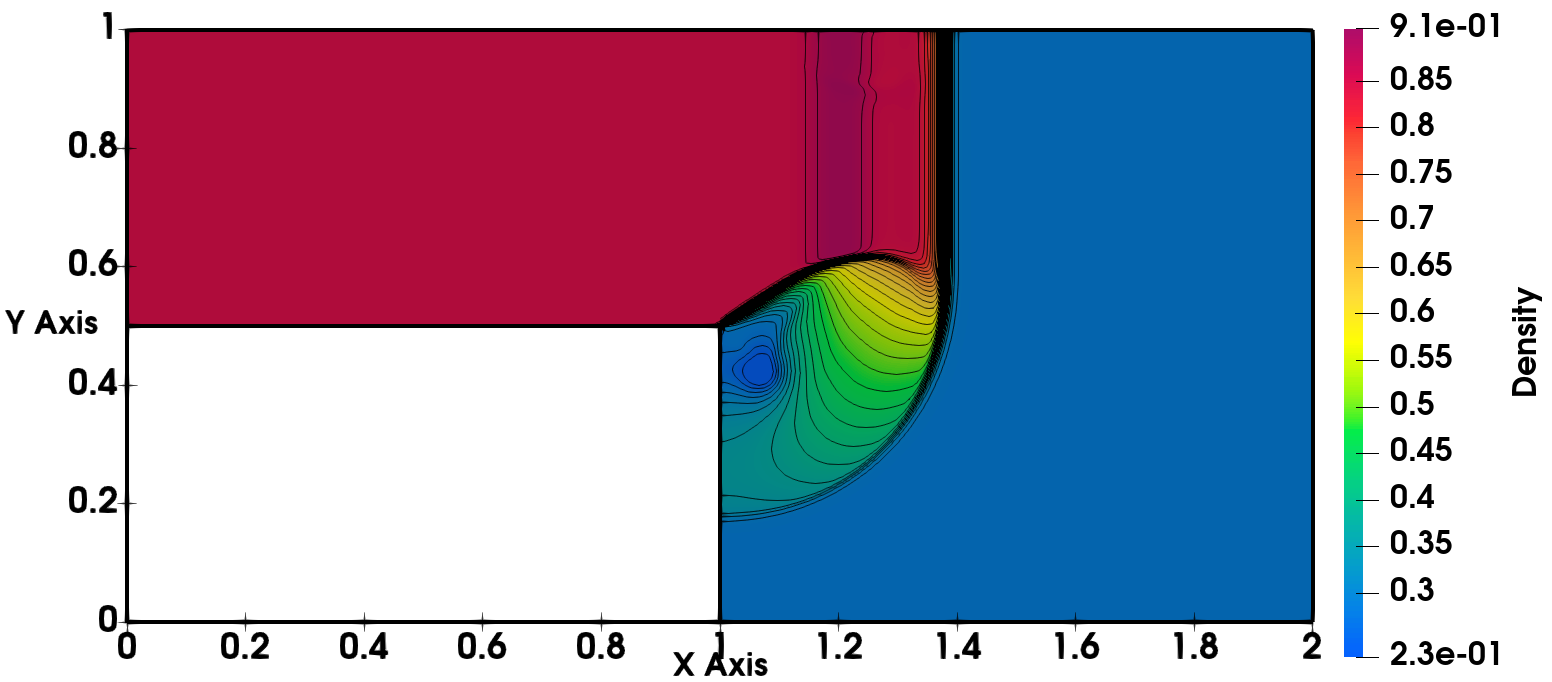}
\caption{Density Contours}
 \label{DC_RICCA_500x500}
\end{subfigure}%
\begin{subfigure}[b]{0.6\textwidth}
\centering
    \includegraphics[scale = 0.15,keepaspectratio]{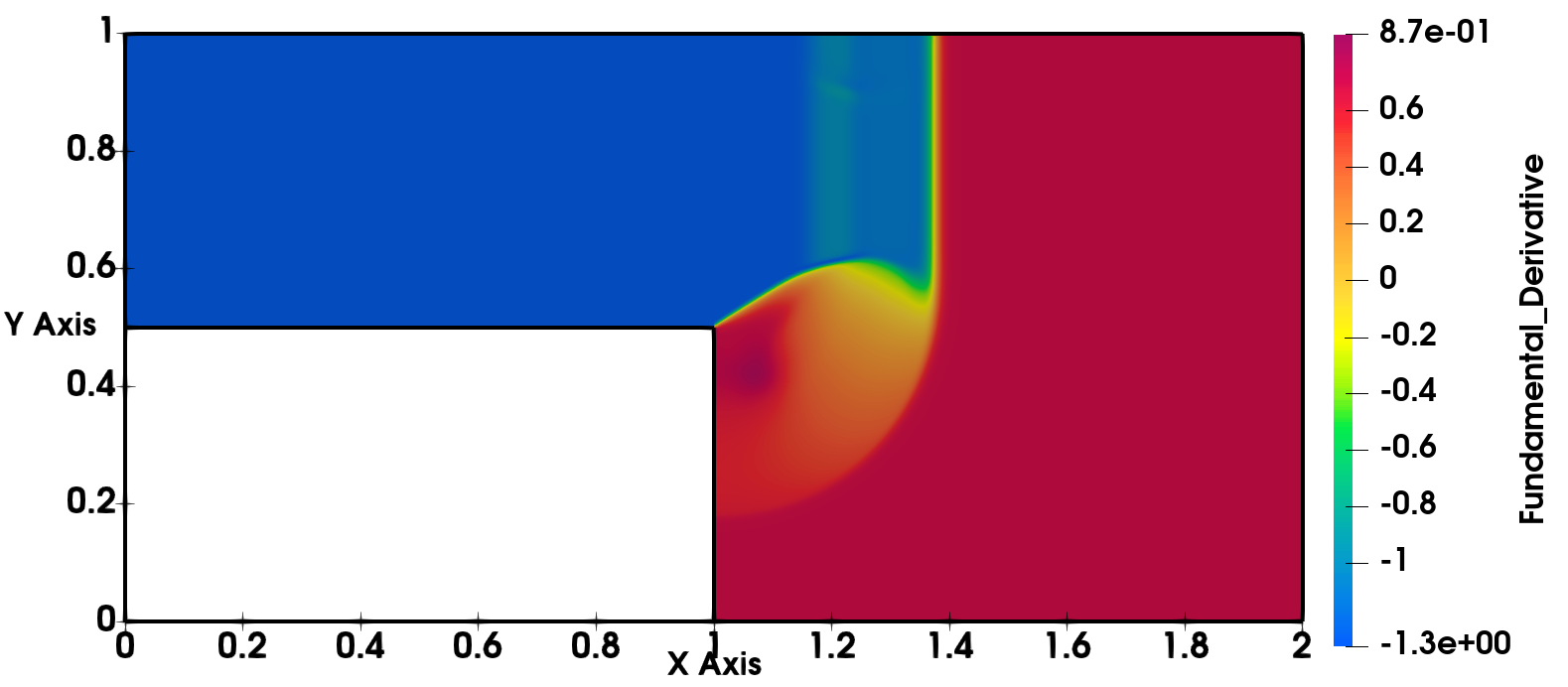}
        \caption{Fundamental Derivative}
 \label{FD_RICCA_500x500}        
\end{subfigure}
}
\caption{Shock Diffraction with $M=1.23$  on $500\times500$ grid using van Der Waals EOS and RICCA at t = 0.4}
 \label{RICCA500x500}
\end{figure}

\begin{figure}[htb!]
\makebox[\textwidth][c]{%
\begin{subfigure}[b]{0.6\textwidth}
\centering
 \includegraphics[scale = 0.15,keepaspectratio]{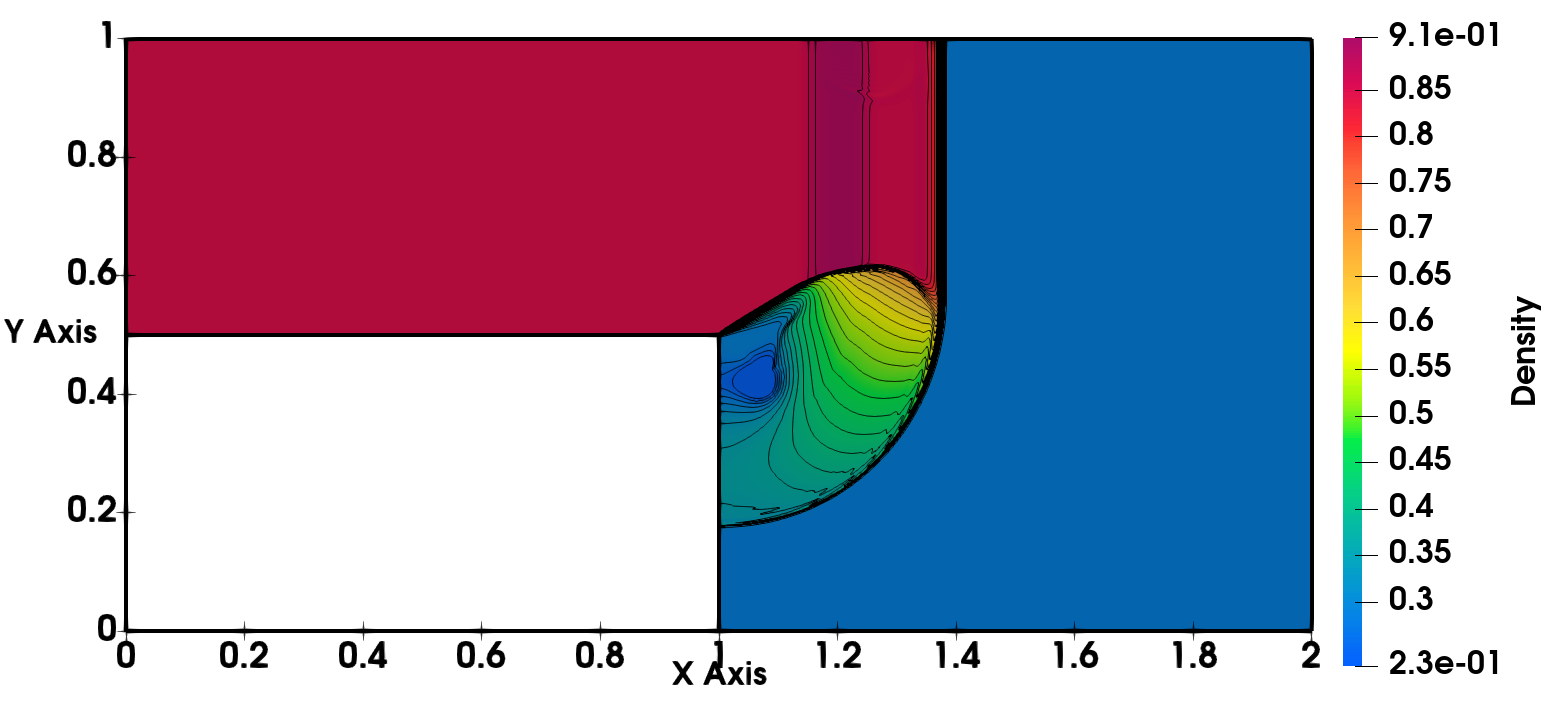}
 \caption{Density Contours}
 \label{DC_Movers+_1000x1000}
\end{subfigure}%
\begin{subfigure}[b]{0.6\textwidth}
\centering
    \includegraphics[scale = 0.15,keepaspectratio]{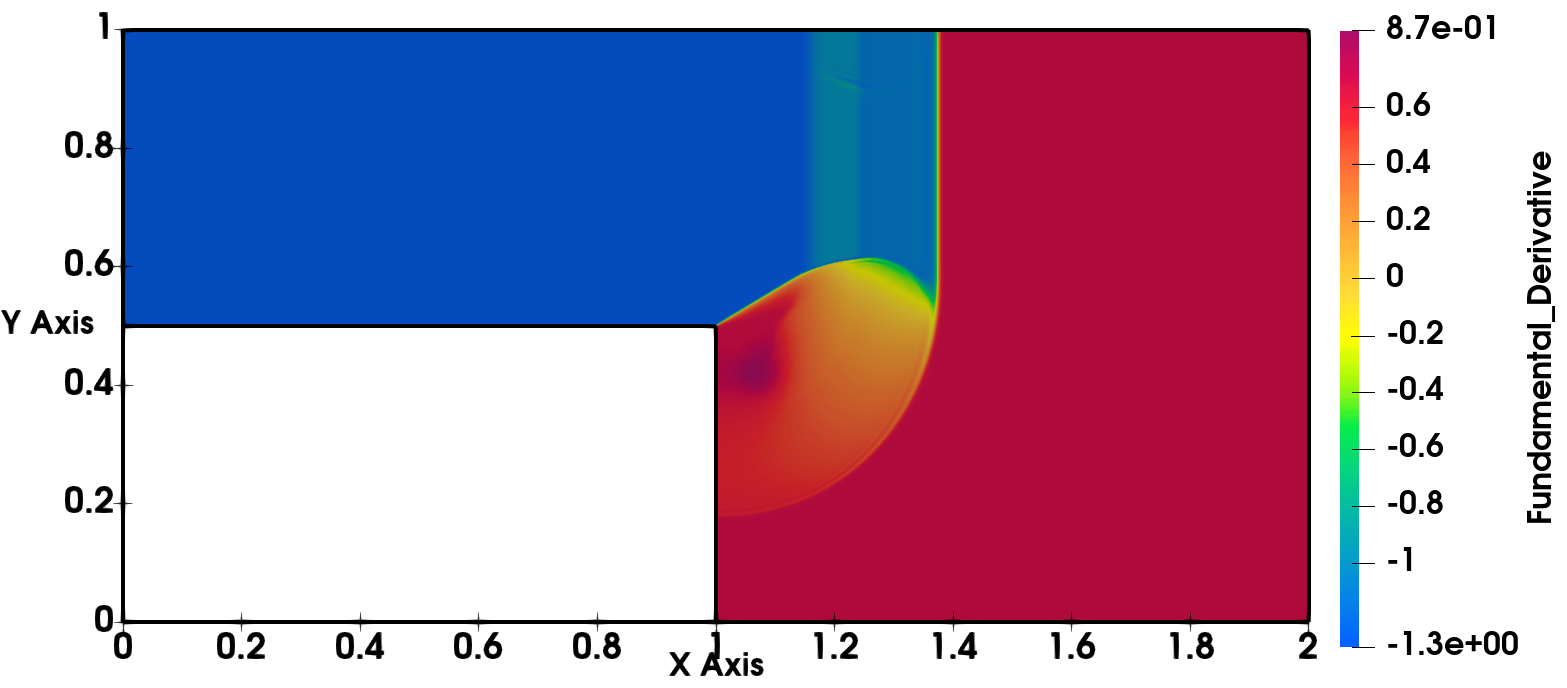}
        \caption{Fundamental Derivative}
 \label{FD_Movers+_1000x1000}        
\end{subfigure}
}
 \caption{Shock Diffraction with $M=1.23$  on $1000\times1000$ grid using van Der Waals EOS and MOVERS + at t = 0.4}
 \label{movers+1000x1000}
 \end{figure}

\begin{figure}[htb!]
\makebox[\textwidth][c]{%
\begin{subfigure}[b]{0.6\textwidth}
\centering
\includegraphics[scale = 0.15,keepaspectratio]{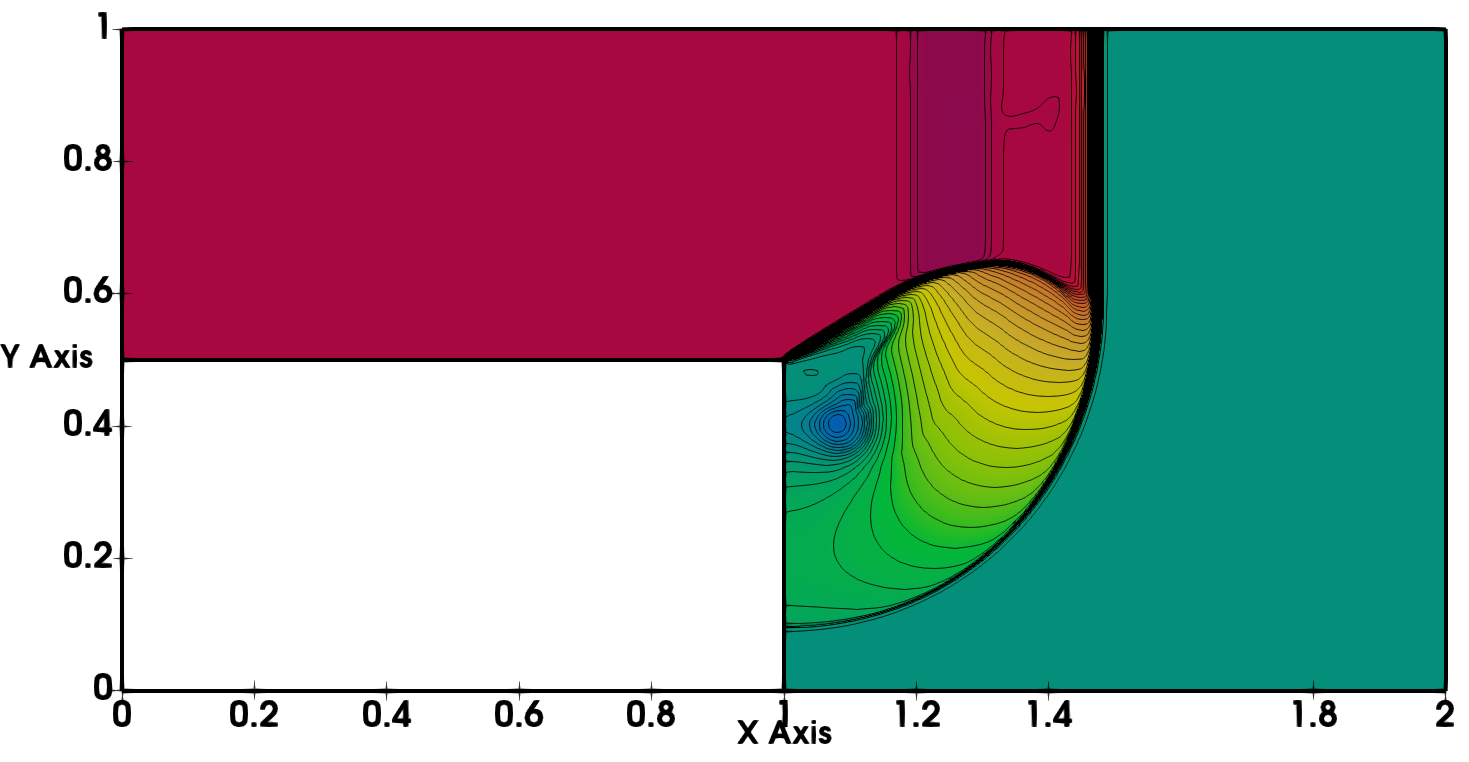}
\caption{Density Contours}
 \label{DC_RICCA_1000x1000}
\end{subfigure}%
\begin{subfigure}[b]{0.6\textwidth}
\centering
    \includegraphics[scale = 0.15,keepaspectratio]{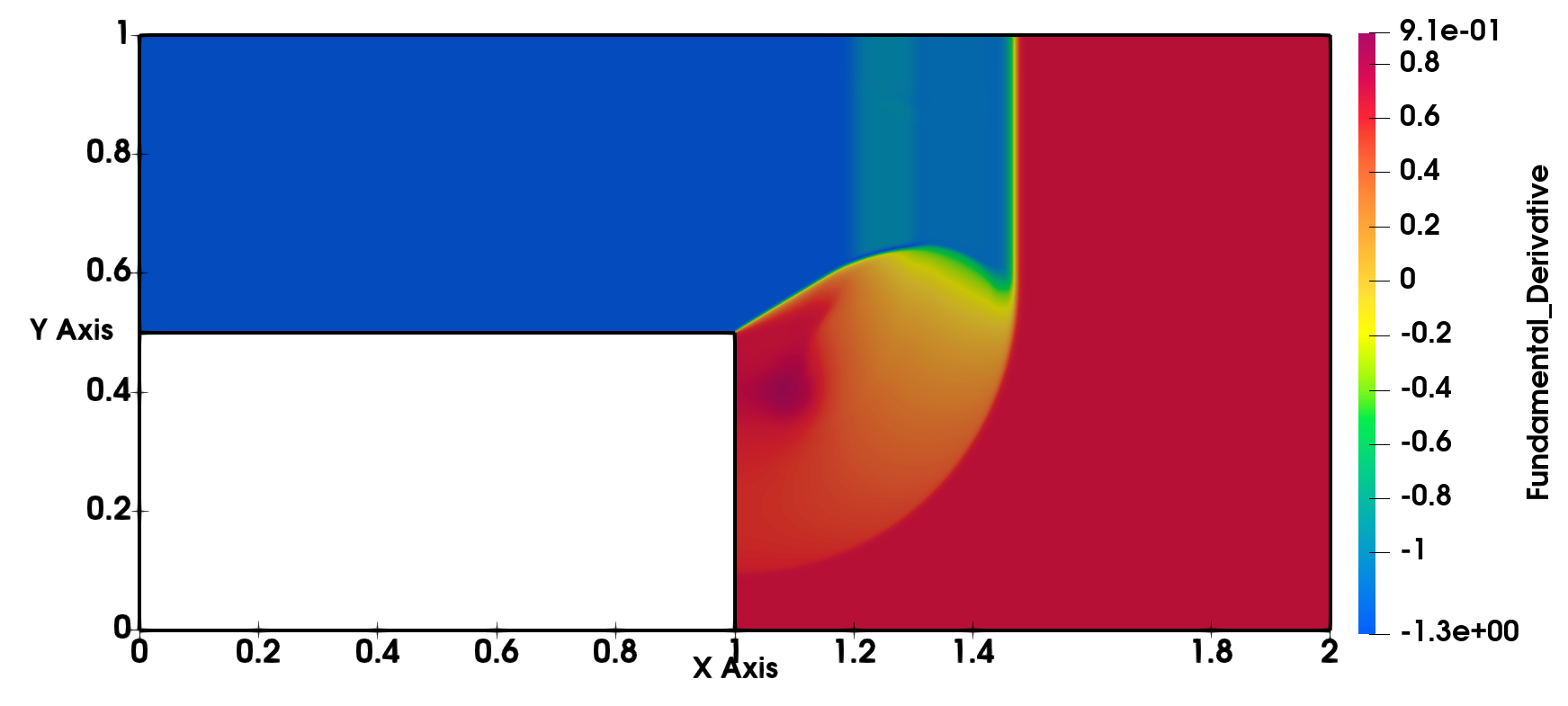}
        \caption{Fundamental Derivative}
 \label{FD_RICCA_1000x1000}        
\end{subfigure}
}
\caption{Shock Diffraction with $M=1.23$  on $1000\times1000$ grid using van Der Waals EOS and RICCA at t = 0.4}
 \label{RICCA1000x1000}
\end{figure}

 \textcolor{black}{As shown in figures (\ref{DC_Movers+_500x500},\ref{DC_RICCA_500x500},\ref{DC_Movers+_1000x1000},\ref{DC_RICCA_1000x1000}) the expansion shock, slip stream and the forward transiting shock are captured well by both the schemes. It can also be observed the change in sign of the fundamental derivative $\Gamma$ from negative to positive as evident from the figures (\ref{FD_Movers+_500x500},\ref{FD_RICCA_500x500},\ref{FD_Movers+_1000x1000},\ref{FD_RICCA_1000x1000}).}

  
 
\section{Conclusions}
Numerical simulations are carried out for dense gases using Euler equations with van der Waals EOS to resolve non-linear non-classical waves, wherein the change in the sign of the fundamental derivative leads to non-classical solutions like physical expansion shocks that satisfy the entropy conditions. Various benchmark test cases in 1D are carried out using MOVERS+ and RICCA. These test cases demonstrate the behaviour of the non-classical waves based on the sign of $\Gamma$. Both the numerical schemes could resolve these waves reasonably well. Further study of wave structures and flow features over simple geometries in 2D are carried out using MOVERS+ and RICCA. Both steady and unsteady test cases are considered in testing the capability of these algorithms. It is observed that the wave fields of dense gas flows are significantly different from those corresponding to the perfect gas. The ability of these algorithms in resolving the flow features with real gas EOS is clearly demonstrated. It can be concluded that these algorithms can be used with any real gas EOS without any modifications for general hyperbolic systems.


%
 \section*{Conflict of interest}
 The authors declare that they have no conflict of interest.

\section*{Data Availability}
 The authors declare that data sharing is not applicable to this article as no new data were created in this study. The data values for the simulation as given in the table (\ref{dgtable:1}) for 1D test cases and    table (\ref{dgtable:2Dtransientcases}, \ref{dgtable:2DSteadystatecases}) for 2D cases are taken from the following references,
 \begin{itemize}
 \item B.P. Brown, B.M. Argrow (1998) , \textit{Nonclassical Dense gas flows for simple geometries}, AIAA Journal \textbf{36(10)},
 \item P. Cinnella (2006), \textit{Roe-Type schemes for dense gas flow computations}, Computers \& Fluids \textbf{35}, p.p. 1264-1281.
 \end{itemize} 




\end{document}